\documentclass{aa}  

\usepackage{natbib}
\bibpunct{(}{)}{;}{a}{}{,}

\usepackage{adjustbox}
\usepackage{graphicx}
\usepackage{longtable}
\usepackage{rotating}
\usepackage{graphics}
\usepackage{psfrag}
\usepackage{amssymb}
\usepackage{xspace}
\usepackage{color}

\usepackage{txfonts}
\usepackage[
bookmarks=true,           
bookmarksnumbered=true,   
colorlinks=true,          
citecolor=blue,           
linkcolor=blue,           
menucolor=blue,           
urlcolor=blue,            
linkbordercolor={0 0 1},  
pdfborder={0 0 1},
frenchlinks=true]{hyperref}

\providecommand{\adsurl}[1]{\href{#1}{ADS}}

\DeclareSymbolFont{UPM}{U}{eur}{m}{n}
\DeclareMathSymbol{\umu}{0}{UPM}{"16}
\let\oldumu=\umu
\renewcommand\umu{\ifmmode\oldumu\else\math{\oldumu}\fi}

\let\oldsim=\sim
\renewcommand\sim{\ifmmode\oldsim\else\math{\oldsim}\fi}
\let\oldpm=\pm
\renewcommand\pm{\ifmmode\oldpm\else\math{\oldpm}\fi}
\newcommand\by{\ifmmode\times\else\math{\times}\fi}

\newbox{\wdbox}
\renewcommand\c{\setbox\wdbox=\hbox{,}\hspace{\wd\wdbox}}
\renewcommand\i{\setbox\wdbox=\hbox{i}\hspace{\wd\wdbox}}

\newcount\timect
\newcount\hourct
\newcount\minct
\newcommand\now{\timect=\time \divide\timect by 60
         \hourct=\timect \multiply\hourct by 60
         \minct=\time \advance\minct by -\hourct
         \number\timect:\ifnum \minct < 10 0\fi\number\minct}

\catcode`@=11

\newcommand\comment[1]{}

\renewcommand\math[1]{$#1$}

\let\atab=&

\let\oldmsp=\sp
\let\oldmsb=\sb
\def\sp#1{\ifmmode
           \oldmsp{#1}%
         \else\strut\raise.85ex\hbox{\scriptsize #1}\fi}
\def\sb#1{\ifmmode
           \oldmsb{#1}%
         \else\strut\raise-.54ex\hbox{\scriptsize #1}\fi}
\newbox\@sp
\newbox\@sb
\def\sbp#1#2{\ifmmode%
           \oldmsb{#1}\oldmsp{#2}%
         \else
           \setbox\@sb=\hbox{\sb{#1}}%
           \setbox\@sp=\hbox{\sp{#2}}%
           \rlap{\copy\@sb}\copy\@sp
           \ifdim \wd\@sb >\wd\@sp
             \hskip -\wd\@sp \hskip \wd\@sb
           \fi
        \fi}
\def\msp#1{\ifmmode
           \oldmsp{#1}
         \else \math{\oldmsp{#1}}\fi}
\def\msb#1{\ifmmode
           \oldmsb{#1}
         \else \math{\oldmsb{#1}}\fi}

\newcommand\Teq{\ifmmode{T\sb{\rm eq}}\else$T$\sb{eq}\fi}
\newcommand\mjup{\ifmmode{M\sb{\rm Jup}}\else$M$\sb{Jup}\fi}
\newcommand\rjup{\ifmmode{R\sb{\rm Jup}}\else$R$\sb{Jup}\fi}
\newcommand\msun{\ifmmode{M\sb{\odot}}\else$M\sb{\odot}$\fi}
\newcommand\rsun{\ifmmode{R\sb{\odot}}\else$R\sb{\odot}$\fi}
\newcommand\mearth{\ifmmode{M\sb{\oplus}}\else$M\sb{\oplus}$\fi}
\newcommand\rearth{\ifmmode{R\sb{\oplus}}\else$R\sb{\oplus}$\fi}
\newcommand\rplanet{\ifmmode{R\sb{\rm p}}\else$R\sb{\rm p}$\fi}
\newcommand\mplanet{\ifmmode{M\sb{\rm p}}\else$M\sb{\rm p}$\fi}
\newcommand\rtransit{\ifmmode{R\sb{\rm t}}\else$R\sb{\rm t}$\fi}

\def\degr{\hbox{$^\circ$}}

\begin{document}

\title{\texttt{PyTranSpot}- A tool for multiband light curve modeling of planetary transits and stellar spots}
\titlerunning{\texttt{PyTranSpot}}

\author{Ines~G.~Juvan\inst{1,2,3} \and
        M.~Lendl\inst{1,4,5} \and
        P.~E.~Cubillos\inst{1} \and
        L.~Fossati\inst{1} \and
        J.~Tregloan-Reed\inst{6} \and
        H.~Lammer\inst{1} \and  
        E.~W.~Guenther\inst{7,8}  \and
        A.~Hanslmeier\inst{2}
}

\institute{Space Research Institute, Austrian Academy of Sciences,
           Schmiedlstrasse 6, A-8042, Graz, Austria \\
           \email{Ines.Juvan@oeaw.ac.at}
           \and
           Institut f\"ur Geophysik, Astrophysik und Meteorologie, Karl-Franzens-Universit\"at, Universit\"atsplatz 5, 8010 Graz, Austria
           \and
            Institut f\"ur Astro- und Teilchenphysik, Universit\"at Innsbruck, Technikerstrasse 25, A-6020 Innsbruck, Austria
           \and
           Observatoire Astronomique de l'Université de Genève, Chemin des Maillettes 51, CH-1290 Sauverny, Switzerland
           \and
           Max Planck Institute for Astronomy, K\"onigstuhl 17, 69117-Heidelberg, Germany
           \and
           Carl Sagan Center, SETI Institute, Mountain View, CA 94043, USA
           \and
           Th\"uringer Landessternwarte Tautenburg, Sternwarte 5, D-07778 Tautenburg, Germany
           \and 
           Visiting scientist Instituto de Astrof\'\i sica de Canarias (IAC), V\'\i a L\'actea s/n, E-38205 La Laguna, Tenerife, Spain
}
\date{Received Month 00, 2017; accepted Month 00, 2017}

\abstract{Several studies have shown that stellar activity features, such as occulted and non-occulted starspots, can affect the measurement of transit parameters biasing studies of transit timing variations and transmission spectra. We present \texttt{PyTranSpot}, which we designed to model multiband transit light curves showing starspot anomalies, inferring both transit and spot parameters. The code follows a pixellation approach to model the star with its corresponding limb darkening, spots, and transiting planet on a two dimensional Cartesian coordinate grid. We combine \texttt{PyTranSpot} with an MCMC framework to study and derive exoplanet transmission spectra, which provides statistically robust values for the physical properties and uncertainties of a transiting star-planet system. We validate \texttt{PyTranSpot}'s performance by analyzing eleven synthetic light curves of four different star-planet systems and 20 transit light curves of the well-studied WASP-41b system. We also investigate the impact of starspots on transit parameters and derive wavelength dependent transit depth values for WASP-41b covering a range of 6200-9200 $\AA$, indicating a flat transmission spectrum.}


\maketitle

\section{Introduction}
\label{introduction}


To date, over 2600\footnote[1]{http://exoplanet.eu/catalog/, \citep{Schneider11}} exoplanets have been confirmed and detected by means of the transit method\footnote[2]{A more detailed description of the transit method can be found in \citet{Winn2010}.}. This method is based on the measurement of a periodic dimming in a stellar light curve, caused by a transiting exoplanet (TEP) passing in front of its host star. By fitting a model to a transit light curve (TLC), one obtains orbital and photometric parameters of the star-planet system, such as the planetary period $P_{\rm{orb}}$, the orbital incination $i$, and the planet-to-star radius ratio \citep[e.g.,][]{Char2000, Winn2010}. In combination with radial velocity or TTV measurements, it is possible to determine the mass of the planet, hence, the average density. Multi-wavelength transmission spectroscopy measurements can constrain the planet's atmospheric properties by comparing the wavelength dependent variations in the planetary radii with theoretical model atmospheres \citep[e.g.,][]{Charb2002, Agol,fortney2008, madhu2009,MRicci2009}.

As already thoroughly discussed \citep[e.g.,][]{SO2011,Ballerini2012,Osh2013,Barros2013, TR2013, TR2015}, the determination of planetary parameters can often be challenging due to the presence of stellar activity features in photometric data sets. Starspots, features which are cooler and thus darker than the surrounding stellar photosphere, can introduce anomalies (``bumps'') in a TLC when they are occulted by a transiting planet \citep[e.g.,][]{Silva2003,Pont2007,Rabus2009,Winn2010b}. The improper treatment of starspots in a TLC fitting process can thus lead to an incorrect determination of the depth, duration and timing of the transit. Additional effects can be introduced by unocculted spots or bright features of stellar activity such as faculae or plages \citep[e.g.,][]{Czesla2009, Desert2011occ, Kipp2012}. The impact of these features on the light curve depends on their location relative to the planetary path across the stellar disk. Unlike occulted starspots, activity features, located in the non-eclipsed area of the stellar surface, do not cause distinct anomalies in a TLC, but have an impact on the overall level of the light curve \citep{Czesla2009}.

Although spots represent sources of noise in a TLC, they can also be seen as useful features to obtain additional information on the observed star-planet system \citep{Kipp2012}. By modeling spots in TLCs, one can constrain the following properties: The latitudinal stellar rotation period $P_{\star}$ which yields a value for the stellar latitudinal rotational velocity, from which it is then possible to infer the stars' age and activity level \citep[e.g., the gyrochronology relationship][]{Barnes2007}\footnote[3]{Note, however, that star-planet interactions may influence gyrochronology age estimations \citep[e.g.,][]{FM2015}.}. Furthermore, one can obtain parameters such as the sky-projected spin-orbit alignment $\lambda$, and, together with an estimate of the inclination angle of the stellar rotation axis and the stellar rotational velocity, the true obliquity of the system can be derived. The determination of the true obliquity then also provides clues on the dynamical evolution of the system \citep[e.g.,][]{Fab2009, Winn2010a, Nutz2011, SO2011a, Des2011}. 

So far, several teams have developed routines to model transit light curves in the presence of starspots:
The \texttt{spotrod} routine \citep{Bek2014} uses a semi-analytical approach to model the TLC of a spotted star, whereas other authors mainly use numerical methods for their astrophysical models: \texttt{SOAP-T} \citep{Osh2013SOAP}, \texttt{PRISM} \citep{TR2013,TR2015}, \texttt{KSint} \citep{Mont2014}, \texttt{ellc} \citep{Maxted2016} and \texttt{StarSim} \citep{Herrero}. Relatable programs have also been developed within the binary star community, such as the \texttt{Wilson-Devinney} (WD) code (\citet{Wilson1971, Wilson1979,Wilson1990,Wilson2008,WD2012}, and references therein) and \texttt{PHEOBE} \citep{Prsa2005, Prsa2016}.

Motivated by the large number of available multicolor photometric transit observations, we developed \texttt{PyTranSpot}\footnote[4]{Researchers interested in \texttt{PyTranSpot} should contact the author.}, a tool which allows for the \textit{simultaneous analysis of transit light curves in the presence of stellar activity and correlated noise}. 
To perform simultaneous analyses of TLCs, derived from various instruments and in different wavelength bands, we combined \texttt{PyTranSpot} with an MCMC framework designed for the determination of exoplanet transmission spectra \citep[][]{Lendlcode}.

This paper is structured as follows: Section~\ref{sec:two} introduces \texttt{PyTranSpot}. Section~\ref{sec:valid} presents the validation of the code using four synthetic star-planet systems. Section \ref{sec:four} describes the analysis and results of 20 WASP-41b transit light curves. Section \ref{sec:five} presents the summary and conclusions of our study.

\section{Modeling Transits and Starspots with PyTranSpot}\label{sec:two}

\subsection{Astrophysical Model and Geometry}
\label{sec:geo}

\begin{figure}
        \centering
	\includegraphics[width=\columnwidth]{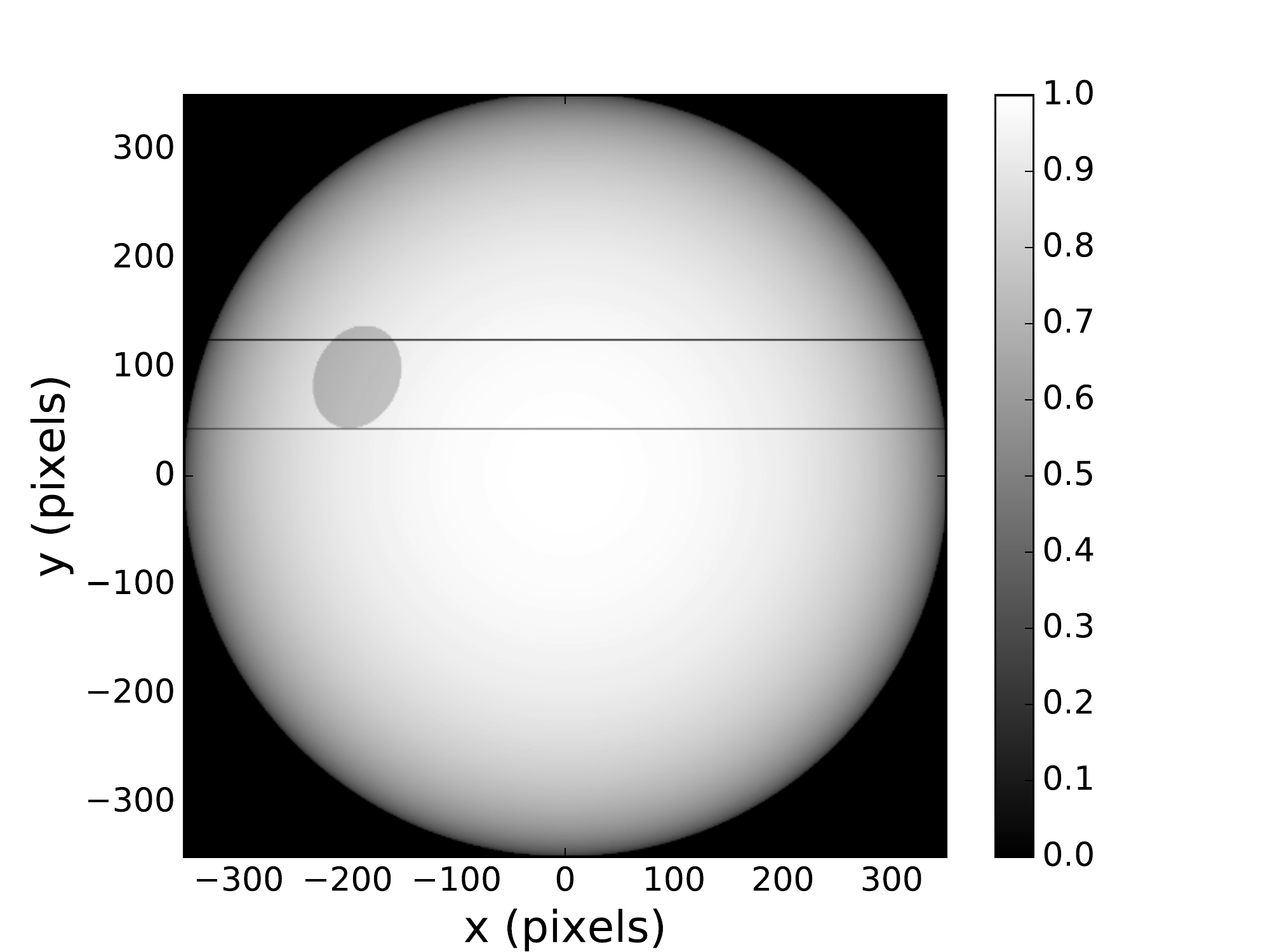}
    \caption{Projection of the stellar sphere with its respective limb darkening, the location of the spot, and the transit cord on the two dimensional grid in Cartesian coordinates. The colorbar on the right hand side indicates the intensity of the stellar flux. Stellar activity features on the stellar sphere are assumed to be homogeneous and circular.}
    \label{fig:map}
\end{figure}

\texttt{PyTranSpot} is based on a pixellation approach, similar to the one used in the \texttt{PRISM} code \citep{TR2013,TR2015}. Within this approach, the pixels are defined as squares on a two-dimensional grid, on which the stellar sphere (and its respective limb darkening), the transit cord, and spots are projected (see Fig.~\ref{fig:map}). The pixel size of the star is determined through dividing the a-priori defined planetary pixel radius by the planet-to-star radius ratio. We assume dark and bright features of stellar activity to be homogeneous and circular over the stellar surface, and they deform to ellipses as they approach the stellar limb. From solar observations we know that sunspots can also appear as a complex group of multiple spots with differing sizes. However, the quality of currently obtained transit light curves is in general not high enough to detect such fine structures. Furthermore, we assume that the stellar rotation period is much longer than the transit.

We calculate the transit light cuve model using:
\begin{equation}
\begin{gathered}
    \Delta F = \frac{F_{\rm{out}}-F_{\rm{transit}}}{F_{\rm{out}}},
	\label{eq:flux}
\end{gathered}
\end{equation}
where $\Delta F$ represents the flux measurement for every timestamp (in and out of transit). $F_{\rm{out}}$ corresponds to the total out of transit flux, also taking into account stellar limb darkening and apparent starspots. From this, we subtract $F_{\rm{transit}}$, which describes the fraction of flux on the stellar sphere occulted by the transiting planet. To derive a normalized light curve, we divide $F_{\rm{out}} - F_{\rm{transit}}$ through $F_{\rm{out}}$. It is also possible to multiply $\Delta F$ with a selected photometric baseline model.

\begin{figure}
        \centering
	\includegraphics[clip, trim=0.5cm 1.0cm 3.5cm 9.5cm, width=0.8\columnwidth]{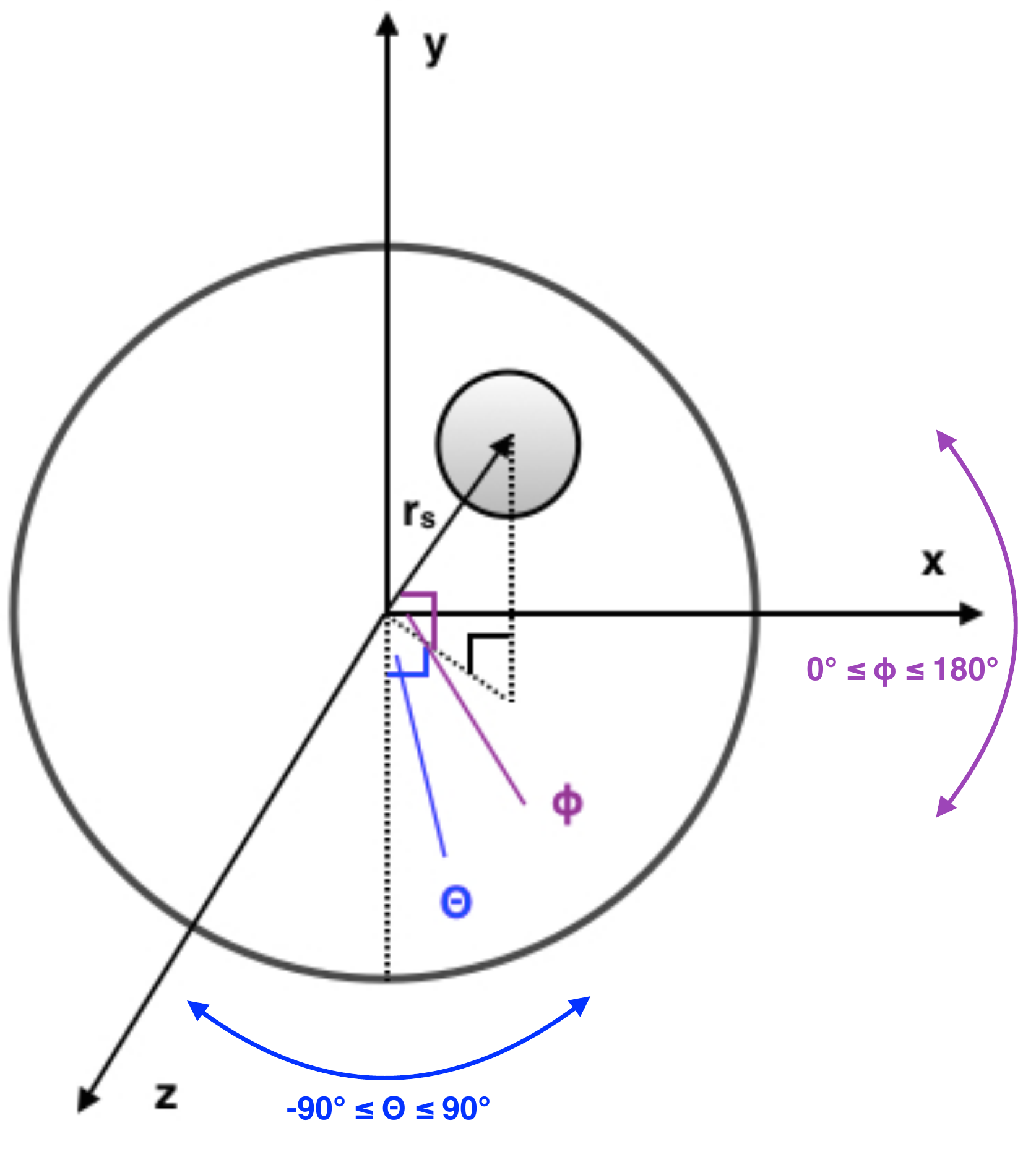}
    \caption{Geometry of the stellar sphere and spot feature. The origin of the coordinate system is the center of the star and spot parameters are defined using spherical coordinates. The grey circle on the stellar surface represents a spot at the distance $r_{\rm{s}}$ (here: stellar radius in \textit{pixels}) from the origin. The longitude $\theta$ (blue) can have values between $-90\degr$ and $90\degr$, whereas $0\degr$ represents the center of the stellar disk. The co-latitude $\phi$ (purple) is defined between $0\degr$ to $180\degr$, with the equator set at $90\degr$, as seen from an observer lying far along the z-axis.}
    \label{fig:spotgeo1}
\end{figure}

\begin{figure}
        \centering
	\includegraphics[clip, trim=0.0cm 13.0cm 5cm 2.7cm, width=0.9\columnwidth]{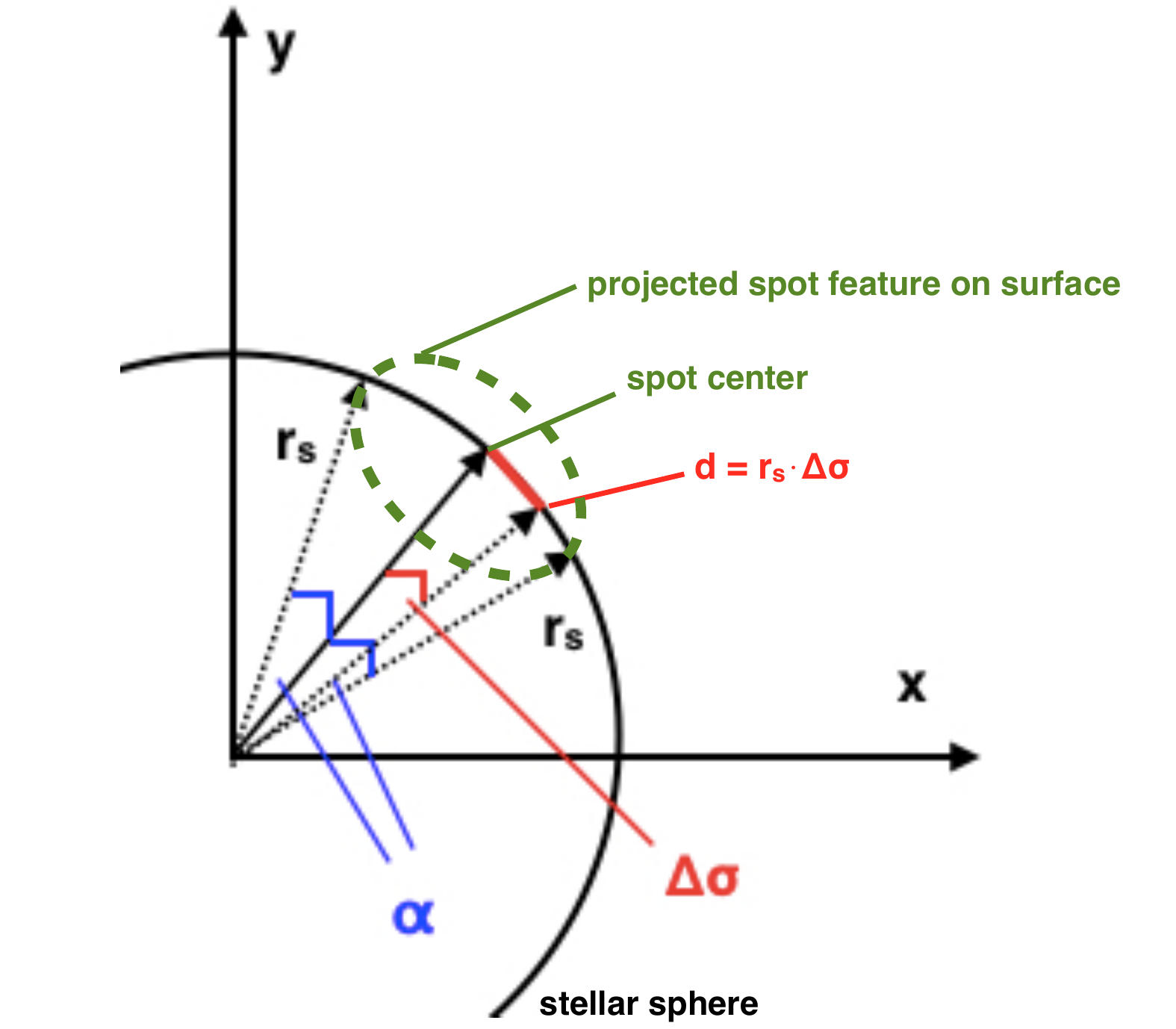}
    \caption{Two-dimensional cut through the stellar sphere, at the position of the spotcenter in the direction perpendicular to the line of sight. The green dashed line represents the spot feature as it would be seen in three dimensions on the stellar sphere. The size of the spot is determined by the angle $\alpha$ (blue). The minor arc $d$ (red) between the spotcenter and any pixel location on the spot can be found through the angle $\Delta\sigma$ (red) times the stellar radius in pixels $r_{\rm{s}}$. Values of $\Delta\sigma$ are derived from Equation \ref{eq:slaw}. Every point on the stellar sphere, corresponding to the spot feature, needs to fulfill the condition: $\Delta\sigma \le \alpha$.}
    \label{fig:spotgeo2}
\end{figure}

Fig.~\ref{fig:spotgeo1} shows the geometry used within our model. The center of the stellar sphere is located at the origin of the three dimensional spherical coordinate system. \texttt{PyTranSpot} does not take into account a fractional area correction, as used within the \texttt{WD} code \citep{Wilson1971, Wilson1979,Wilson1990,Wilson2008,WD2012}. We argue that this effect is negligible, as the resulting loss of accuracy is much smaller than the noise currently present in observations. However, to obtain precise photometric transit and spot parameters, we recommend to use a planetary pixel radius between 15 and 50 pixels \citep{TR2015}. On the stellar sphere, every point is described by the two angles (longitude $\theta$, co-latitude $\phi$) and the distance to the stellar center (stellar radius $r_{\rm{s}}$ in pixels). The longitude $\theta$ varies between $-90\degr$ and $90\degr$, with the center of the stellar disk corresponding to a value of $0\degr$. The co-latitude $\phi$ ranges from $0\degr$ to $180\degr$, with the stellar equator set at $90\degr$. 
Fig.~\ref{fig:spotgeo2} illustrates the projection of a spot on to the stellar sphere, as seen from a two dimensional perspective. The observer is assumed to lie far along the z-axis. To determine the pixels on the stellar sphere, which correspond to the starspot, we implement the following boundary condition: If the angle $\Delta\sigma$ between the pixel on the sphere and the spotcenter is greater than the angular radius of the spot $\alpha$, then this pixel does not belong to the spot. The values for $\Delta\sigma$ are derived by using the spherical law of cosines:
\begin{equation}
    \cos(\Delta\sigma) = \cos(\phi_{\rm{spot}}) \cdot \cos(\phi) + \sin(\phi_{\rm{spot}}) \cdot \sin(\phi) \cdot \cos(\Delta\theta),
	\label{eq:slaw}
\end{equation}
where $\phi_{\rm{spot}}$ and $\phi$ are the co-latitudes of the spotcenter and the surrounding pixels, respectively. The value $\Delta \theta$ represents the absolute difference in longitude between the spotcenter and the pixel center.

\subsection{Model parameters}
\label{sec:input}
\texttt{PyTranSpot} calculates the transit and spot model by using the following photometric and orbital parameters:

\begin{itemize}
\item phase offset from the orbital phase = 0.0, which indicates the transit midpoint;
\item planet-to-star radius ratio $r_{\rm{P}}/r_{\rm{S}}$;
\item orbital inclination $i$ in degrees;
\item semi-major axis $a$ in units of the stellar radius $a/R_{\rm{S}}$;
\item planetary orbital period ${P}_{\rm{orb}}$ in days;
\item orbital eccentricity $e$;
\item argument of periastron $\omega$ in degrees;
\item linear $u_{1}$ and quadratic $u_{2}$ limb darkening coefficients;
\item coefficients of the baseline model functions.
\end{itemize}
To model one or more spots on the stellar sphere, each starspot is characterized by the following set of parameters:
\begin{itemize}
\item spot longitude $\theta$ in degrees ($-90 \degr \leqslant \theta \leqslant 90 \degr $);
\item spot co-latitude $\phi$ in degrees ($0 \degr$ to $180 \degr$);
\item spot size $\alpha$ in degrees;
\item spot contrast $\rho_{\rm{spot}}$ ($\rho_{\rm{spot}}\in[0,1]$\footnote[5]{This limitation corresponds to the modeling of (un)occulted dark spots. When modeling bright features (e.g., faculae or plages), the contrast can take values of $\rho_{\rm{spot}} > 1.0$. However, defining the contrast of, e.g., a plage region is more complex as it depends
on its location on the stellar sphere and the stellar type \citep[e.g.,][]{Beeck48B, Beeck49B, Thaler2014}.}, where $1$ equals the surrounding stellar photosphere).
\end{itemize}

\subsection{Treatment of stellar limb darkening}
\label{sec:noise}

\texttt{PyTranSpot} employs the quadratic limb darkening law \citep[][LDL]{Kopal}, which is the most commonly used LDL in TLC analyses. The quadratic LDL describes the specific intensity of a star $I$($\mu$) on the surface as
\begin{equation}
    I(\mu)/I(0) = 1 - u_{1}(1-\mu) - u_{2}(1-\mu)^2,
	\label{eq:LD}
\end{equation}
where $I(0)$ defines the intensity at the center of the stellar disk, $\mu$ is the cosine of the angle between the line of sight of the observer and the unit normal to the stellar surface, and $u_{1}$ and $u_{2}$ are the quadratic limb darkening coefficients (LDCs).

We make this choice because, unlike three- or four-parameter LDLs, the quadratic two-parameter law preserves the curvature in the intensity of the star, without trying to model bumps due to starspots in the light curve \citep{Csiz2013}. Also, using a two parameter law reduces the number of free parameters in the model \citep{Kipp2012}, which is especially important when analyzing large data sets which require a great number of MCMC jump parameters. To guarantee physical values of the quadratic LDCs $u_{1}$ and $u_{2}$, we implement the following conditions proposed by \citet{Kipp2013} within our astrophysical model:
\begin{equation}
\begin{aligned}
    u_{1} + u_{2} & < 1, \\
    u_{1} & > 0, \\
    u_{1} + 2u_{2} & > 0 .
	\label{eq:kipp1}
\end{aligned}
\end{equation}

These conditions make sure that the intensity profile remains everywhere-positive, and guarantees a monotonically decreasing intensity profile from the center of the stellar sphere to the limb.

\subsection{Determination of Transmission Spectra}\label{sec:wave} 

We use \texttt{PyTranSpot} in combination with the MCMC framework developed by \citet[][]{Lendlcode}. This MCMC framework employs the statistical package \texttt{MCCubed} (see Section \ref{sec:mcmc}) and allows for the simultaneous analysis of multiband transit observations yielding filter dependent planet-to-star radius ratios $r_{\rm{P}}/r_{\rm{S}}$. When calculating a transmission spectrum using multiband TLCs, the light curves share the same model parameters (see Section \ref{sec:input}). Exceptions are the filter dependent limb darkening coefficients, the baseline model coefficients, and the spot parameters. These parameters are fitted individually. To derive the multiband planet-to-star radius ratios, we fit a wavelength dependent offset to a reference $r_{\rm{P}}/r_{\rm{S}}$ value \citep[][]{Lendlcode}.


\section{Code Validation using Synthetic Light Curves}\label{sec:valid}

\subsection{Synthetic Data Sets}
As a first approach to test the performance of \texttt{PyTranSpot}, we create four synthetic star-planet systems yielding a total of eleven light curves. The first system SYNTH-1 (Table \ref{tab:synth1}) consists of a hot Jupiter planet with a solar-like host star. We derive one transit light curve without stellar activity features (SYNTH-1a). The star-planet system SYNTH-2 (Table \ref{tab:synth2}) hosts a Saturn-size planet orbiting an active, solar-like star. The generated light curve (SYNTH-2a) shows a starspot crossing at the limb of the star. The third system SYNTH-3 is similar to SYNTH-2, but with a shorter orbital period, and we create five synthetic observations of SYNTH-3. The five light curves of this system are referred to as SYNTH-3a, -3b, -3c, -3d, and -3e. We further create one of the light curves (SYNTH-3e) with an anomaly due to an occulted starspot. Table \ref{tab:synth3} shows the system properties and the spot parameters which refer to the SYNTH-3e light curve. System SYNTH-4 (Table \ref{tab:synth4}) also describes a hot Jupiter planet orbiting a solar-like star. For this system, we simulate a simultaneous observation of the same transit event, measured in the Johnson B,V, and Cousins R, I filters. We further assume that the corresponding $r_{\rm{P}}/r_{\rm{S}}$ value has no wavelength dependence. The resulting light curves SYNTH-4a to SYNTH-4d show the same occulted starspot. We calculate the wavelength dependent spot contrasts using equation (1) of \citet{Silva2003}, using a blackbody approximation, a solar-like effective temperature for the host star of $T_{\rm{eff}}$ = 5772 K, and a spot temperature of $T_{\rm{spot}}$ =  4772 K. We generate all simulated TLCs using \texttt{PyTranSpot} and add Gaussian noise to the calculated flux.

\renewcommand{\arraystretch}{1.2}
\begin{table}
	\centering
	\caption{Photometric properties of the synthetic light curve SYNTH-1a. This light curve is for a spot-free transit.}\label{tab:synth1}
	\begin{tabular}{p{5cm}lcr} 
		\hline
		Parameter and Unit & Symbol & SYNTH-1 \\
		\hline
		Ratio of fractional radii  & $r_{\rm{P}}/r_{\rm{S}}$ & $0.113$ \\
                  Orbital inclination (\degr) & $i$ & $87.18$ \\
		Relative semi-major axis & $a/R_{\rm{S}}$ & $7.88$ \\
                  Planetary orbital period (d) & ${P}_{\rm{orb}}$ & $2.788$ \\
                  Transit mid time (HJD-2450000) & $T_{\rm{mid}}$ & $5817.70461$ \\
                  Orbital eccentricity & $e$ & $0.0$ \\
                  Argument of Periastron (\degr) & $\omega$ & $0.0$ \\
                  Linear LDC & $u_{\rm{1}}$ &  $0.5$\\
                  Quadratic LDC &  $u_{\rm{2}}$ & $0.2$ \\
                  Added Random Noise Level (ppm) & & 700  \\
		\hline
	\end{tabular}
\end{table}

\renewcommand{\arraystretch}{1.2}
\begin{table}
	\centering
	\caption{Photometric properties of the synthetic light curve SYNTH-2a. This TLC shows one occulted starspot during the transit.}\label{tab:synth2}
	\begin{tabular}{p{5cm}lcr} 
		\hline
		Parameter and Unit & Symbol & SYNTH-2 \\
		\hline
		Ratio of fractional radii  & $r_{\rm{P}}/r_{\rm{S}}$ & $0.0694$ \\
                  Orbital inclination (\degr) & $i$ & $88.6$ \\
		Relative semi-major axis & $a/R_{\rm{S}}$ & $11.8$ \\
                  Planetary orbital period (d) & ${P}_{\rm{orb}}$ & $4.2$ \\
                  Transit mid time (HJD-2450000) & $T_{\rm{mid}}$ & $54129.722$ \\
                  Orbital eccentricity & $e$ & $0.0$ \\
                  Argument of Periastron (\degr) & $\omega$ & $0.0$ \\
                  Linear LDC & $u_{\rm{1}}$ &  $0.646$\\
                  Quadratic LDC &  $u_{\rm{2}}$ & $0.048$ \\
                  Added Random Noise Level (ppm) & & 400 \\
		\hline
		Spot No.1 Parameter and Unit & Symbol  &  \\
		\hline
                  Longitude (\degr) & $\theta$ & $-56.0$ \\
                  Co-Latitude (\degr) & $\phi$ & $75.0$\\
                  Size (\degr) & $\alpha$ &$15.0$\\
                  Contrast & $\rho_{\rm{spot}}$ & $0.78$\\
		\hline
	\end{tabular}
\end{table}

\renewcommand{\arraystretch}{1.2}
\begin{table}
	\centering
	\caption{Photometric properties of the synthetic system SYNTH-3. One of the five TLCs (SYNTH-3e) shows an occulted starspot during transit.}\label{tab:synth3}
	\begin{tabular}{p{5cm}lcr}
		\hline
		Parameter and Unit & Symbol & SYNTH-3 \\
		\hline
		Ratio of fractional radii  & $r_{\rm{P}}/r_{\rm{S}}$ & $0.0694$ \\
                  Orbital inclination (\degr) & $i$ & $88.6$ \\
		Relative semi-major axis & $a/R_{\rm{S}}$ & $11.8$ \\
                  Planetary orbital period (d) & ${P}_{\rm{orb}}$ & $2.2$ \\
                  Transit mid time (HJD-2400000) & $T_{\rm{mid}}$ & $55433.421$ \\
                  Orbital eccentricity & $e$ & $0.0$ \\
                  Argument of Periastron (\degr) & $\omega$ & $0.0$ \\
                  Linear LDC & $u_{\rm{1}}$ &  $0.646$\\
                  Quadratic LDC &  $u_{\rm{2}}$ & $0.048$ \\
                  Added Random Noise Level (ppm) & & 400 \\
		\hline
		Spot No.1 Parameter and Unit & Symbol  &  \\
		\hline
                  Longitude (\degr) & $\theta$ & $-30.0$ \\
                  Co-Latitude (\degr) & $\phi$ & $73.0$\\
                  Size (\degr) & $\alpha$ &$11.0$\\
                  Contrast & $\rho_{\rm{spot}}$ & $0.77$\\
		\hline
	\end{tabular}
\end{table}

\renewcommand{\arraystretch}{1.2}
\begin{table}
	\centering
	\caption{Photometric properties of the synthetic system SYNTH-4. The five TLCs show the same transit event and occulted starspot observed in the Johnson B, V, and Cousins R, I filters.}\label{tab:synth4}
	\begin{tabular}{p{5cm}lcr}
		\hline
		Parameter and Unit & Symbol & SYNTH-4 \\
		\hline
		Ratio of fractional radii  & $r_{\rm{P}}/r_{\rm{S}}$ & $0.092$ \\
                  Orbital inclination (\degr) & $i$ & $89.1$ \\
		Relative semi-major axis & $a/R_{\rm{S}}$ & $11.7$ \\
                  Planetary orbital period (d) & ${P}_{\rm{orb}}$ & $5.72$ \\
                  Transit mid time (HJD-2400000) & $T_{\rm{mid}}$ & $55197.4130$ \\
                  Orbital eccentricity & $e$ & $0.0$ \\
                  Argument of Periastron (\degr) & $\omega$ & $0.0$ \\
                  Spot longitude (\degr) & $\theta$ & $-10.0$ \\
                  Spot co-Latitude (\degr) & $\phi$ & $74.0$\\
                  Spot size (\degr) & $\alpha$ &$5.0$\\
                  Added Random Noise Level (ppm) & & 300 \\
	         \hline
                  SYNTH-4a (Johnson B filter): &  & \\
                  Linear LDC & $u_{\rm{1}}$ &  $0.6328$\\
                  Quadratic LDC &  $u_{\rm{2}}$ & $0.1834$ \\
                  Spot contrast & $\rho_{\rm{spot}}$ & $0.301$\\
		\hline
                  SYNTH-4b (Johnson V filter):&  & \\
                  Linear LDC & $u_{\rm{1}}$ &  $0.4306$\\
                  Quadratic LDC &  $u_{\rm{2}}$ & $0.2995$ \\
                  Spot contrast & $\rho_{\rm{spot}}$ & $0.383 $\\
		\hline
                  SYNTH-4c (Cousins R filter):&  & \\
                  Linear LDC & $u_{\rm{1}}$ &  $0.3316$\\
                  Quadratic LDC &  $u_{\rm{2}}$ & $0.3275$ \\
                  Spot contrast & $\rho_{\rm{spot}}$ & $0.434 $\\
		\hline
                  SYNTH-4d (Cousins I filter):&  & \\
                  Linear LDC & $u_{\rm{1}}$ &  $0.2486$\\
                  Quadratic LDC &  $u_{\rm{2}}$ & $0.3288$ \\
                  Spot contrast & $\rho_{\rm{spot}}$ & $0.537 $\\
		\hline
	\end{tabular}
\end{table}

\subsection{Light Curve Analysis}\label{sec:synthanalysis}

We analyze the synthetic light curves of the systems SYNTH-1 and SYNTH-2 individually, whereas the transit light curves SYNTH-3a to SYNTH-3e, and SYNTH-4a to SYNTH-4d are analyzed simultaneously. For each MCMC analysis, we run 10 chains with a total of 600\,000 samples. Only for the simultaneous analysis of SYNTH-4a to SYNTH-4d, we use 23 chains with a total of 2\,000\,000 samples due to the larger number of free paramteres. The MCMC jump parameters are those listed in Section \ref{sec:input}, except for the planetary orbital period, the eccentricity, and the argument of periastron, which are fixed. We also do not consider photometric baseline models during these analyses. Furthermore, we analyze the spotted TLCs (SYNTH-2a and SYNTH-3e) assuming a spot-free transit model to investigate the impact of starspots on transit parameters. The simultaneous multiband observations of SYNTH-4 are used to study the ability of \texttt{PyTranSpot} to reproduce the transit and spot parameters while fitting for the filter dependent limb darkening coefficients and spot contrasts.

We recalculate errorbars of each data set using uncorrelated (white) and correlated (red) noise factors \citep{Winn08,Gillon2010}. This guarantees that uncertainties are not being underestimated in the course of the analysis. We perform the analysis for each system multiple times (at least ten repetitions for SYNTH-1 and SYNTH-2, and three for SYNTH-3 and SYNTH-4) to make sure that the obtained results are consistent and thus robust.

\subsubsection{Statistical Package}\label{sec:mcmc}

To carry out the statistical analysis, we use the open-source package
Multi-Core Markov-Chain Monte Carlo
\citep[\texttt{MCcubed},][]{Pato2016}. \texttt{MCcubed}\footnote[6]{https://github.com/pcubillos/MCcubed} is a
Python/C code that provides statistically-robust model optimization
via Levenberg-Marquardt minimization and credible-region estimation
via MCMC sampling. \texttt{MCcubed} assesses the goodness-of-fit between the
model and data through $\chi\sp{2}$ statistics, considering
user-defined flat or Gaussian priors.
To sample the parameter space, we choose the Snooker Differential-Evolution algorithm
\citep{Braak2008SnookerDEMC}, which
automatically adjusts the scale and orientation of the proposal
distribution. The code checks for MCMC convergence through the 
\citet{GelmanRubin1992} statistics.

\subsection{Results}

\subsubsection{Transit and Spot Parameters}

\begin{figure}
        \centering
	\includegraphics[width=0.85\columnwidth]{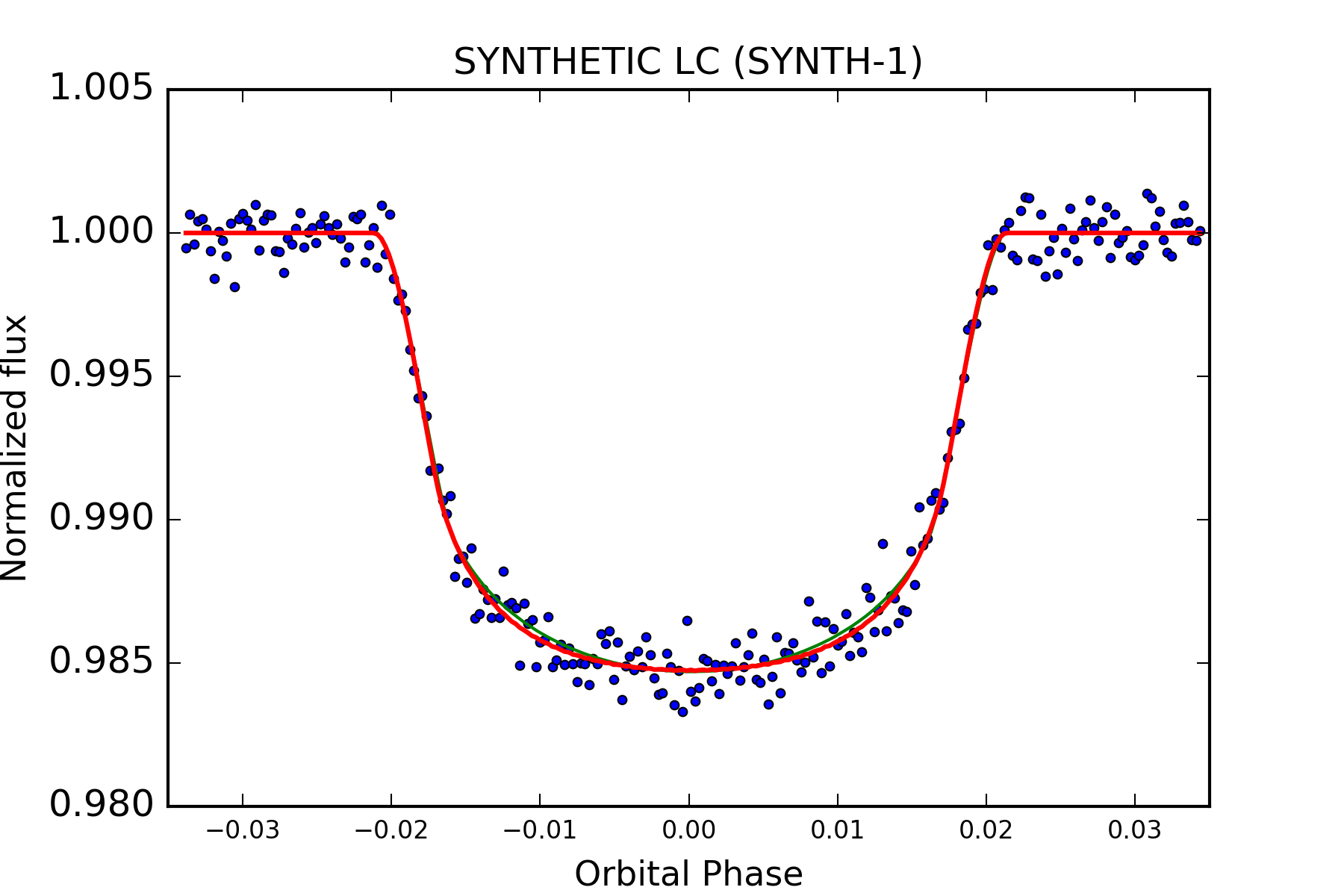}
    \caption{Transit light curve (blue dots) of the synthetic star-planet system SYNTH-1, with the derived best-fit model (red line) and the original light curve model (green line). The obtained photometric parameters can be found in Table \ref{tab:ressynth}. The (O-C) residuals and the difference between the original and best-fit light curve models are presented in Figure \ref{fig:Synth1res}.}
    \label{fig:Synth1}
\end{figure}

\begin{figure}
        \centering
	\includegraphics[width=0.85\columnwidth]{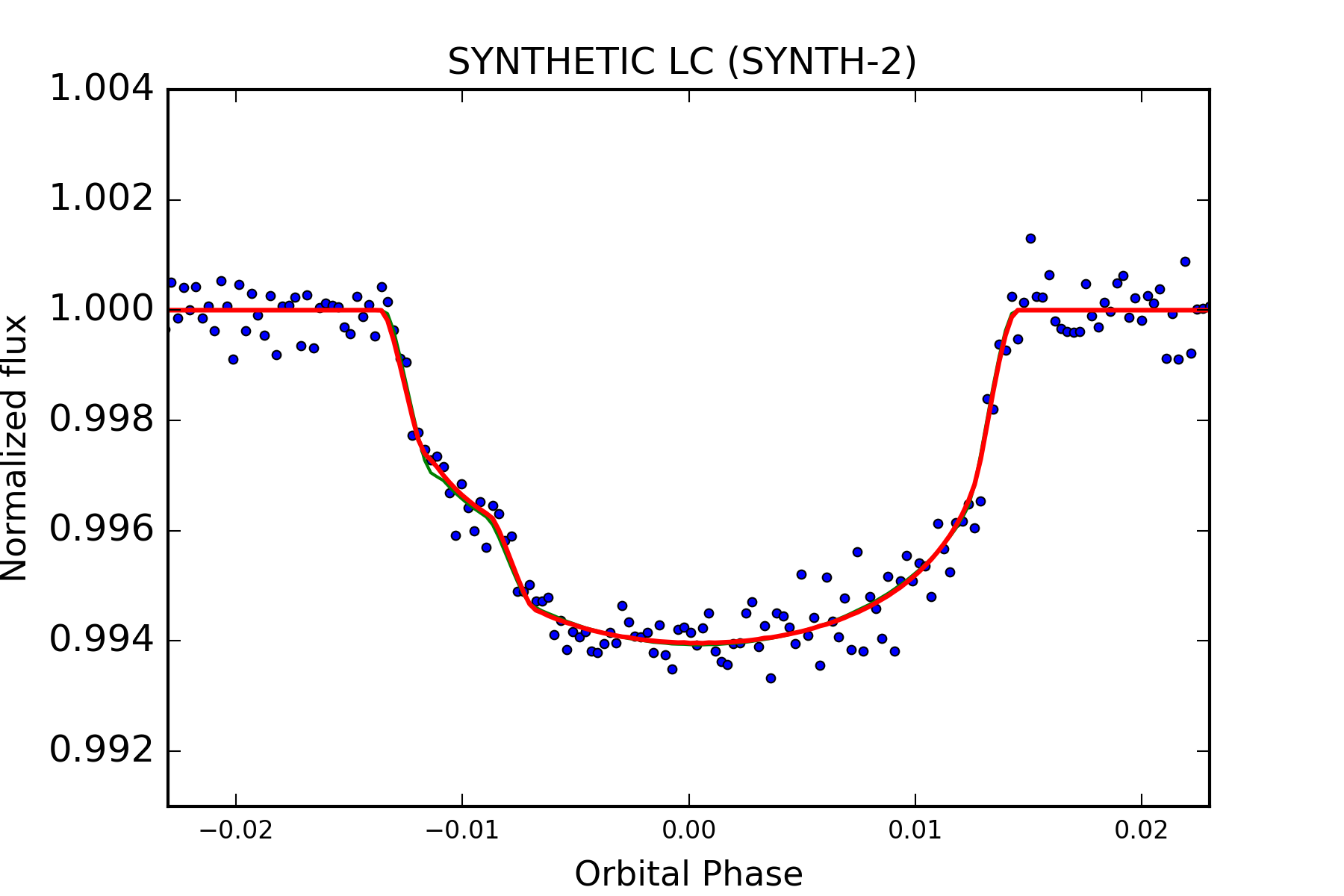}
    \caption{Synthetic light curve SYNTH-2a (blue dots), with the derived best-fit model (red line) and the original light curve model (green line). This synthetic light curve shows a starspot crossing at the limb of the star around phase -0.01. The obtained photometric properties and spot parameters can be found in Tables \ref{tab:ressynth} and \ref{tab:spotsyn}. The (O-C) residuals and the difference between the original and best-fit TLC models are shown in Figure \ref{fig:Synth2res}.}
    \label{fig:Synth2}
\end{figure}

\begin{figure}
        \centering
	\includegraphics[clip, trim=0.0cm 0.7cm 0.0cm 0.7cm, width=0.85\columnwidth]{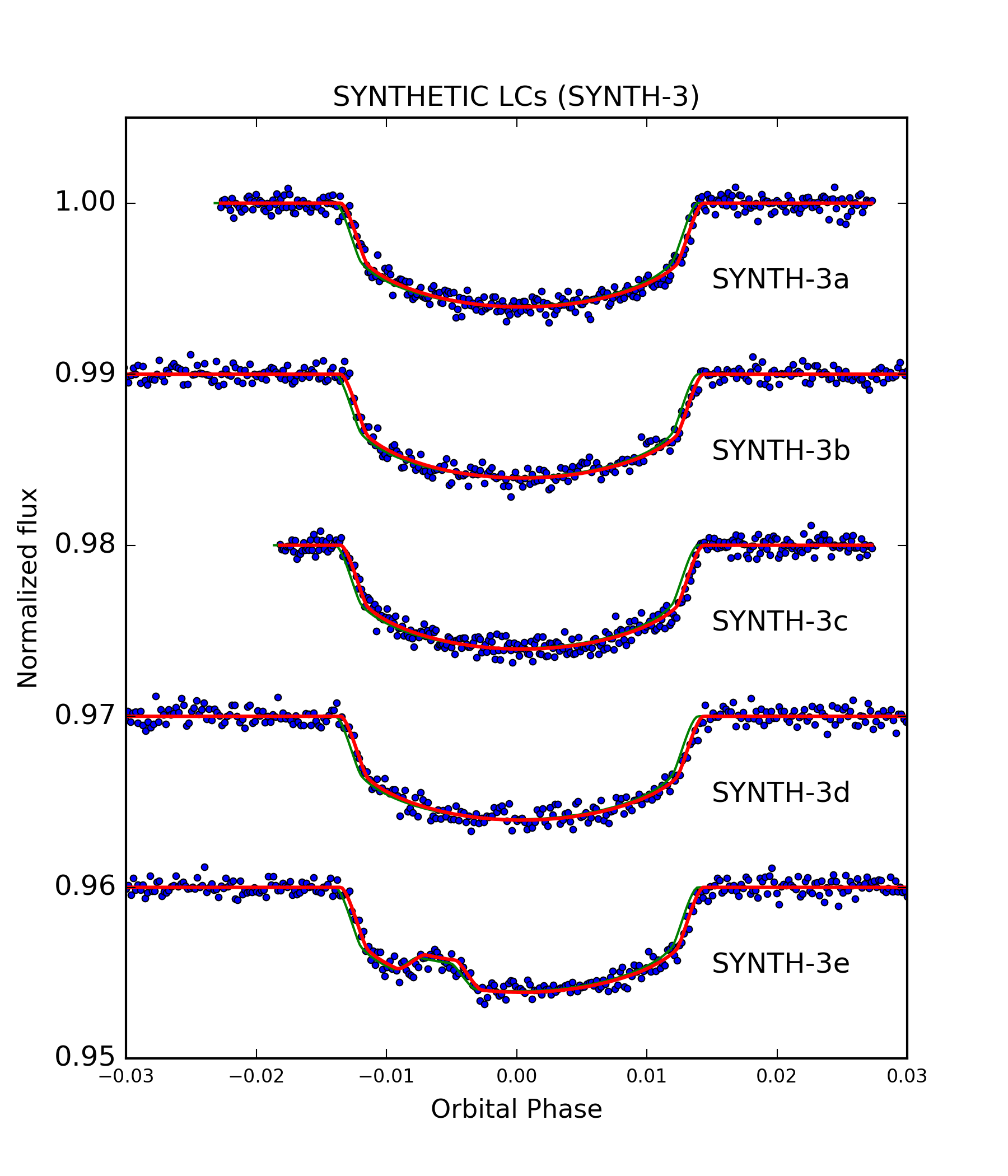}
    \caption{Synthetic light curves SYNTH-3a to SYNTH-3e, with the derived best-fit models (red line) and the original light curve model (green line). Light curve SYNTH-3e is showing a starspot anomaly located close to the limb of the star around phase -0.005. The obtained photometric properties and spot parameters can be found in Tables \ref{tab:ressynth} and \ref{tab:spotsyn}. The (O-C) residuals and the difference between the original and best-fit TLC models are shown in Figures \ref{fig:Synth3res} and \ref{fig:Synth3mod}, respectively.}
    \label{fig:Synth3}
\end{figure}

\begin{figure}
        \centering
	\includegraphics[clip, trim=0.0cm 0.7cm 0.0cm 0.7cm, width=0.85\columnwidth]{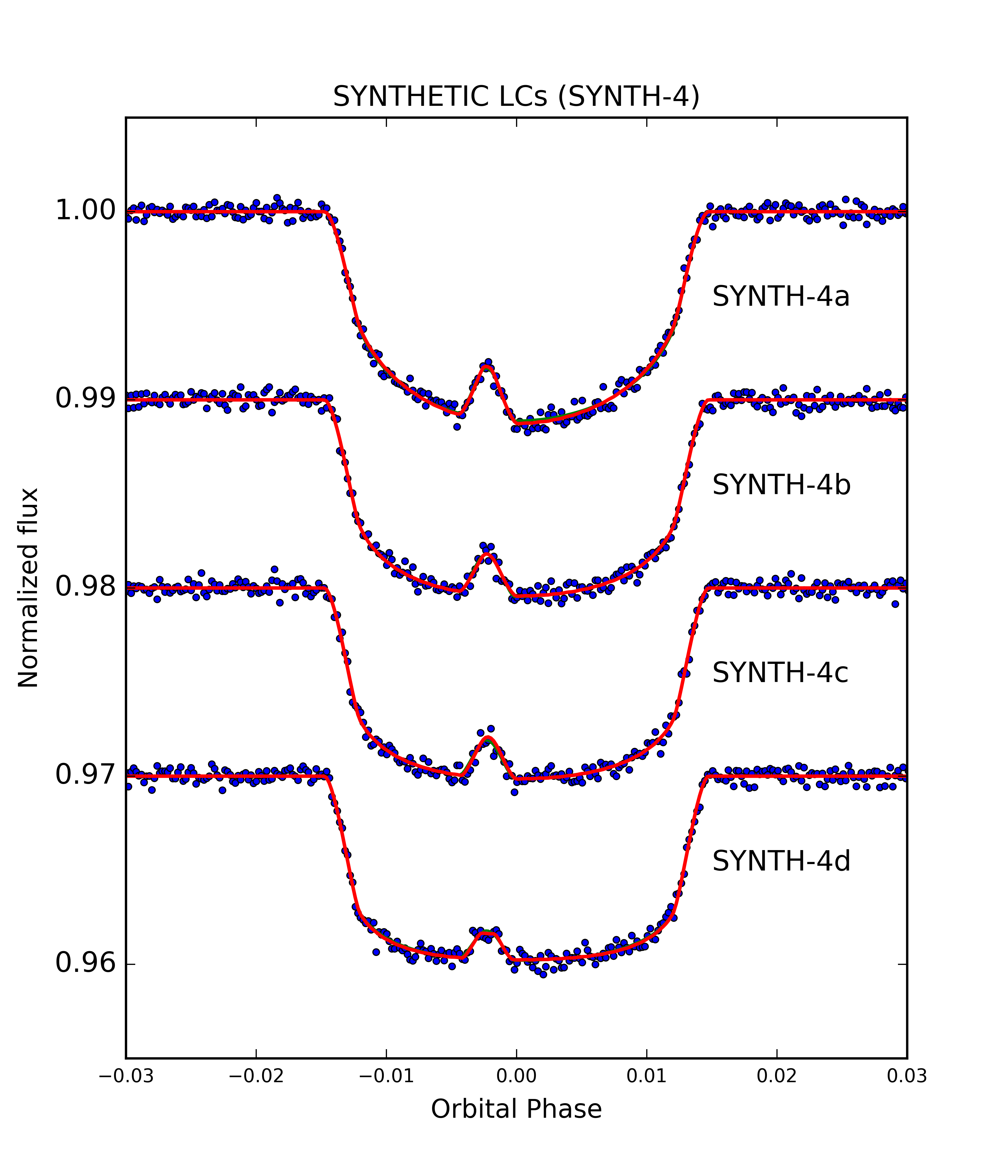}
    \caption{Synthetic light curves SYNTH-4a to SYNTH-4d, with the derived best-fit models (red line) and the original light curve model (green line). The TLCs all show the same transit event and occulted starspot around phase -0.002. The obtained photometric and spot parameters can be found in Tables \ref{tab:ressynth4} and \ref{tab:spotsyn4}. The (O-C) residuals and the difference between the original and best-fit TLC models are shown in Figures \ref{fig:Synth4res} and \ref{fig:Synth4mod}, respectively.}
    \label{fig:Synth4}
\end{figure}

We present in Figures \ref{fig:Synth1}, \ref{fig:Synth2}, \ref{fig:Synth3}, and \ref{fig:Synth4} the synthetic light curves with the resulting best-fit models. Figures \ref{fig:Synth1res}, \ref{fig:Synth2res}, \ref{fig:Synth3res}, \ref{fig:Synth3mod}, \ref{fig:Synth4res}, and \ref{fig:Synth4mod} show the (O-C) residuals and a visualization of the differences between the input- and best-fit light curve models. Tables \ref{tab:ressynth}, \ref{tab:spotsyn},  \ref{tab:ressynth4}, and \ref{tab:spotsyn4} give the derived transit and spot properties. A comparison of the obtained parameters of systems SYNTH-1, SYNTH-2 and SYNTH-3 with their original system parameters (Tables \ref{tab:synth1}, \ref{tab:synth2}, and \ref{tab:synth3}) shows that we can recover the majority of the input values consistently within one sigma. Only the LDCs of SYNTH-1a slightliy differ from the original input values, but the differences are smaller than 1.3$\sigma$. We also find that the derived results of the repeated MCMC analysis for each system yield consistent values. The transit parameters, spot locations, filter dependent limb darkening coefficients (LDCs) and spot contrasts of our multiband SYNTH-4 TLCs could also be reproduced within 1.2$\sigma$. An exception is the derived spot contrast for light curve SYNTH-4a (B filter), which differs by 2$\sigma$ from the original value. However, using the derived spot contrasts, a stellar effective temperature of $T_{\rm{eff}}$ = 5772 K, and equation (1) of \citet{Silva2003}, we derive an average spot temperature of $T_{\rm{spot}}$ =  $4956^{+245}_{-175}$ K. This value agrees within 1.1$\sigma$ with the original spot temperature of $T_{\rm{spot}}$ = 4772 K. The ability of our code to reproduce also the wavelength dependent spot contrasts is an important result as simultaneous multiband observations of starspots can help to contrain starspot temperatures. Without such simultaneous measurements, the spot temperature remains strongly correlated with its radius \citep[e.g.,][]{TR2013, Mancini2014}.

We also perform a test to calculate the theoretically expected errorbar on the planet-to-star radius ratio to compare it with the one derived from our analysis. For this, we calculate the expected uncertainty on $r_{\rm{P}}/r_{\rm{S}}$ of SYNTH-1. We assume that our data is only affected by white noise and obtain a theoretical uncertainty $\sigma_{r_{\rm{P}}/r_{\rm{S}}}$ of $0.00052$. We find that our derived uncertainty is $\gtrsim 3$ times larger than the theoretical one. This is a reasonable result as we assume the transit shape to be a simple trapezoid when calculating the expected uncertainty, but in practice, the TLC model is more complex (i.e., more free parameters). In addition, to quantify the effect of a lower pixel resolution on the derived parameters and uncertainties, we re-analyze the TLCs SYNTH-1 and SYNTH-2. We perform identical analyzes, but vary the planetary pixel radius $r_{P}=10, 15, 20$ from the original one $r_{P}=50$. We find that for all cases, the derived values using a lower pixel resolution differ by  $\leq 0.1 \sigma$ from the results using $r_{P}=50$. We also calculate the differences between the low resolution and the $r_{P}=50$ light curve models and derive rms values between 8 - 20 ppm, which are much lower than the noise in SYNTH-1 (700 ppm) and SYNTH-2 (400 ppm). This agrees with \citet{TR2015}, who also find that a higher pixel resolution does not significantly increase the numerical resolution of the resulting parameters and uncertainties. However, the use of $r_{P}=50$ is recommended, as a smaller planet pixel radius does affect the smoothness of the resulting best-fit model.

\subsubsection{Impact of Starspots on Transit Parameters}
We compare the transit parameters, which we derive from the analyses of the spotted TLCs SYNTH-2a and SYNTH-3e (assuming both a spot and spot-free model), and find the following: The majority of the transit parameters agree well within 1.4$\sigma$. We also identify that the phaseoffset value, which we derive from SYNTH-2a's spot-free model, differs by 2.5$\sigma$. Not taking into account the spot feature, which is located at the limb of SYNTH-2, seems to affect the determination of the transit duration and hence, the transit midtime. The effect of starspots on the measured transit duration and timing confirms findings by various authors \citep[e.g.,][]{Silva2010, SO2011, Osh2013}. 

\section{WASP-41b: a broadband 6200-9200 $\AA$ transmission spectrum in the presence of starspots}\label{sec:four}
To further test the performance of \texttt{PyTranSpot}, we use data of the well-studied WASP-41 system. WASP-41 is one of the targets of the Wide Angle Search for Planets project (\citet{Poll2006}, WASP\footnote[7]{http://wasp-planet.net}), and is a V=11.6 G8V star which is known to show magnetic activity and rotational modulation on a period of $18.41\pm 0.05$ days \citep{Maxted2011}. The systems' transiting hot Jupiter, WASP-41b, has a measured planetary mass and radius of $0.94$ $M_{\rm{Jup}}$ and $1.06$ $R_{\rm{Jup}}$, respectively. The study of \citet{SW2016} on WASP-41b discusses that some of the transit light curves show anomalies in brightness due to occulted spots. Table \ref{tab:wasp41} presents the derived spot parameters of \citet{SW2016}. From modeling two spot features, and assuming that these are caused by the same spot, the authors determined the rotation period of the host star to be $P_{\star}=18.6 \pm 1.5$ days, and a sky-projected orbital obliquity of $\lambda = 6 \pm 11\degr$. Since the host star is magnetically active showing TLCs with and without starspots, WASP-41b represents an ideal object to further test our routine. Our aim is to reproduce the results of \citet{SW2016}, to investigate the effect of starspots on the transit parameters, and to derive a broadband transmission spectrum in the range of 6200-9200 $\AA$ for WASP-41b. To accomplish this, we use archival data in different wavelength bands together with yet unpublished transit light curves of WASP-41b.

\begin{table}
	\centering
	\caption{Photometric and starspot properties of the WASP-41 system, as taken from \citet{SW2016}.}\label{tab:wasp41}
	\begin{tabular}{lcr} 
		\hline
		Parameter and Unit & Symbol & WASP-41 \\
		\hline
		Stellar mass ($M_{\rm{\sun}}$) & $M_{\rm{\star}}$ & $0.987 \pm 0.021$ \\
		Stellar radius ($R_{\rm{\sun}}$) & $R_{\rm{\star}}$ & $0.866 \pm 0.009$ \\
                  Age (Gyr) & $ $ & ${1.2}^{+1.0}_{-0.0}$ \\
		Effective Temperature (K) & $T_{\rm{eff}}$ & $5546 \pm 50$ \\
                  Orbital semi-major axis (AU) & $a$ & $0.0410 \pm 0.0003$ \\
                  Planetary mass ($M_{\rm{Jup}}$) & $M_{\rm{b}}$ & $0.977 \pm 0.020$ \\
                  Planetary radius ($R_{\rm{Jup}}$) & $R_{\rm{b}}$ & $1.178 \pm 0.015$ \\
                  Planetary surface gravity ($m{s}^{-2}$) & $g_{\rm{b}}$ & $17.45 \pm 0.46$ \\
                  Planetary density ($\rho_{\rm{Jup}}$) & $\rho_{\rm{b}}$ & $0.558 \pm 0.020$ \\
                  Equilibrium Temperature (K) & $T_{\rm{eq}}$ & $1242 \pm 12$ \\
                  Sky-projected obliquity ($\degr$) & $\lambda$  & $6 \pm 11$ \\
		\hline
		Spot Parameters and Units & Symbol & LC 2015/05/13  \\
		\hline
                  Spot No. 1: & & \\
                  Longitude (\degr) & $\theta$  & $-36.3 \pm 4.5$  \\
                  Co-Latitude (\degr) & $\phi$  & $74.7 \pm 10.3$ \\
                  Spot size (\degr) & $\alpha$  & $10.4 \pm 6.5$ \\
                  Spot contrast & $\rho_{\rm{spot}}$  & $0.80 \pm 0.14$ \\
		\hline
		Spot Parameters and Units & Symbol & LC 2015/05/17  \\
		\hline
                  Spot No. 1$\star$: & & \\
                  Longitude (\degr) & $\theta$  & $-13.9 \pm 5.2$ \\
                  Co-Latitude (\degr) & $\phi$  & $61.8 \pm 6.5$ \\
                  Spot size (\degr) & $\alpha$  & $14.3 \pm 3.2$ \\
                  Spot contrast & $\rho_{\rm{spot}}$  & $0.86 \pm 0.08$ \\
                  Spot No. 2: & & \\
                  Longitude (\degr) & $\theta$  & $23.7 \pm 1.6$ \\
                  Co-Latitude (\degr) & $\phi$  & $81.7 \pm 6.5$\\
                  Spot size (\degr) & $\alpha$  & $14.3 \pm 3.2$ \\
                  Spot contrast & $\rho_{\rm{spot}}$  & $0.89 \pm 0.06$\\
		\hline
	\end{tabular}
\tablefoot{$\star$ We note that the spot parameters (Spot No. 1), which were originally presented in \citet{SW2016}, have been corrected and we show the updated value in this Table (J. Southworth; private communication May 2017).} 
\end{table}

\subsection{Data}
\subsubsection{EulerCam observations}

We observed a total of nine transits of WASP-41 between January 2011 and April 2012 with EulerCam, 
the CCD imager installed at the 1.2m Euler telescope at ESO La Silla, Chile. From these nine unplublished transit light curves, three show evidence of occulted stellar spots. The observations were carried out using a r'-Gunn filter and the telescope was slightly defocused to improve efficiency and PSF sampling. Table \ref{tab:Ecam} gives a summary of the individual observations. We reduced the data using aperture photometry and tested a range of apertures and reference stars, selecting those which produce the minimal residual scatter of the fitted transit light curve. Refer to \citet{Lendl2012} for details on the instrument and the data reduction procedures. 

\begin{table}
\centering                        
\caption{\label{tab:Ecam}Observing log of EulerCam observations of WASP-41} 
\begin{tabular}{lccl}
\hline\hline 
Date & airmass & average & exposure \\
(UT) & range & FWHM [arcsec] & time [s] \\
\hline
 2011/01/31 & 1.0 - 1.8 & 2.6  & 60, 80, 160 \\
 2011/04/02 & 1.0 - 2.0 & 2.8  & 120      \\
 2011/05/12 & 1.0 - 1.3 & 2.8  & 100      \\
 2011/05/15 & 1.0 - 1.2 & 2.3  &  50      \\
 2011/05/24 & 1.0 - 2.1 & 3.6  & 120      \\
 2012/03/09 & 1.0 - 1.9 & 3.0  &  70, 90  \\
 2012/03/12 & 1.0 - 1.2 & 2.9  &  80, 90 \\
 2012/03/18 & 1.0 - 1.7 & 3.0  &  80     \\
 2012/04/30 & 1.0 - 1.3 & 3.0  &  60     \\
\hline
\end{tabular}
\end{table}

\subsubsection{DFOSC I observations}

For our study, we also consider a set of four transit light curves from \citet{SW2016}. The authors observed four transits of WASP-41b with the Danish Faint Object Spectrograph and Camera (DFOSC) instrument, which is installed on the 1.54m Danish Telescope at ESO La Silla, Chile. The object was observed using a Bessell I filter. \citet{SW2016} describe the observations and data reduction of these data.
The DFOSC data are of specific interest for this paper as two of the four light curves show occulted starspots. The authors also observed two additional transits of WASP-41b using the 84cm telescope at Observatorio Cerro Amazones in Antofagasta. Due to their lower quality, these two transit light curves are not included in our study.

\subsubsection{TRAPPIST and DFOSC R observations}

\citet{NVM2016} present eight transits of the WASP-41 system from which we adopt the five data sets obtained with TRAPPIST \citep{Gillon2011,Jehin} in the I+z filter, and two transit light curves observed with the DFOSC instrument in the Bessell R filter (see Table \ref{tab:obsinfo}). We note that only one of the DFOSC R light curves covers the full transit. In addition, we decide not to include the TLC observed with the Faulkes Telescope South (FTS) telescope located at Siding Spring Observatory, as the observation was affected by poor weather conditions.

\begin{table*}
	\centering
	\caption{WASP-41b observations analyzed in this work. $N_{\rm{Data}}$ is the number of data set frames and the last two columns indicate the applied photometric model function and the available external parameters of the observations. The sources of the respective light curve data sets can be found below the table.}
	\label{tab:obsinfo}
	\begin{tabular}{lccccr} 
		\hline
		Telescope & Filter & Date of & $\rm{N}_{\rm{Data}}$ & Baseline  & additional \\
                  &         & obs.      &                           &   function     &  Info\\
		\hline
                  EulerCam (1)& r' & 2011/01/31 & 109 & $p(t^{2})+p(xy^{2})$ & $\star$\\
		EulerCam (1)& r' & 2011/04/02  & 103 & $p(t^{2})+p(xy^{2})$ & $\star$\\
		EulerCam (1)& r' & 2011/05/12  & 83 & $p(t^{2})+p(xy^{2})$ & $\star$\\
	         EulerCam (1)& r' & 2011/05/15  & 196 & $p(t^{2})$ & $\star$\\
                  EulerCam (1)& r' & 2011/05/24  & 102 & $p(t^{2})+p(xy^{2})$ & $\star$\\
		EulerCam (1)& r' & 2012/03/09  & 169 & $p(t^{2})+p(xy^{2})$ & $\star$\\
		EulerCam (1)& r' & 2012/03/12  & 141 & $p(t^{2})$ & $\star$\\
	         EulerCam (1)& r' & 2012/03/18  & 155 & $p(t^{2})$ & $\star$\\
                  EulerCam (1)& r' & 2012/04/30  & 189 & $p(t^{2})+p(xy^{2})$ & $\star$\\
                  DFOSC (2)& I & 2014/05/31  & 155 &$p(t^{2})$ & none\\
                  DFOSC (2)& I & 2015/05/10  & 148 &$p(t^{2})$ & none\\
                  DFOSC (2)& I & 2015/05/13  & 159 &$p(t^{2})$ & none\\
                  DFOSC (2)& I & 2015/05/17  & 166 &$p(t^{2})$ & none\\
		TRAPPIST (3)& I+z & 2011/03/21  & 435 & $p(t^{2})+p(FWHM^{2})$& FWHM\\
		TRAPPIST (3)& I+z & 2011/04/02  & 407 & $p(t^{2})+p(FWHM^{2})$ & FWHM\\
		TRAPPIST (3)& I+z & 2011/05/12  & 311 & $p(t^{2})$ & FWHM\\
		TRAPPIST (3)& I+z & 2012/03/09  & 575 & $p(t^{2})+p(FWHM^{2})$ & FWHM\\
		TRAPPIST (3)& I+z & 2013/04/19  & 1158 & $p(t^{2})+p(FWHM^{2})$ & FWHM\\
	         DFOSC (3)& R & 2013/04/19 & 102 & $p(t^{2})$ & none \\
	         DFOSC (3)& R & 2013/04/23 & 83 & $p(t^{2})$ & none \\
		\hline
	\end{tabular}
\tablefoot{(1) These data sets are newly released observations obtained from the 1.2m Euler telescope at ESO La Silla, Chile, (2) \citet{SW2016}, (3) \citet{NVM2016}.
 
$\star$ xshift, yshift, airmass, FWHM, sky}
\end{table*}

\subsection{Light Curve Analysis}\label{sec:ResW41}

We model the WASP-41b transit light curves individually, as well as simultaneously, using \texttt{PyTranSpot} within the MCMC transmission spectroscopy framework \citep[][]{Lendlcode}. For all individual and simultaneous fitting processes, we additionally analyze the spotted TLCs assuming a spot-free model. From the simultaneous analyses, we further derive a transmission spectrum in the range of 6200-9200 $\AA$ for WASP-41b. We remark that we infer only one R-band planet-to-star radius ratio for the transmission spectrum, combining the EulerCam r' and DFOSC R TLCs.

Within the simultaneous MCMC analysis, the TLCs share the same transit parameters (see Section \ref{sec:input}), except for the bandpass dependent limb darkening coefficients. In addition, the spot parameters and baseline coefficients are analyzed separately for each light curve. The wavelength dependent $r_{\rm{P}}/r_{\rm{S}}$ values are derived through fitting an offset to a reference planet-to-star radius ratio. Following \citet{Gillon2010}, the coefficients describing the baseline models are calculated for every MCMC step by applying least-square minimization \citep[][]{Lendlcode}. As discussed in \citet{SW2016}, we also fix the orbital inclination to the value $i = 88.7\degr$, which restricts the strong correlation between the planet's orbital inclination and the spot latitude. Furthermore, we fix the planetary orbital period to ${P}_{\rm{orb}} = 3.05$ days, as well as the eccentricity $e$ and the argument of periastron $\omega$, which are both set to zero. We use the quadratic limb darkening coefficients, which we inferred from \texttt{JKTLD}\footnote[8] {http://www.astro.keele.ac.uk/jkt/codes/jktld.html} \citep{JKTLD}, as starting values for our MCMC analysis. Errorbars for each light curve are rescaled as discussed in Section \ref{sec:synthanalysis}. Each run consists of 10-20 parallel MCMC chains with a total of up to 1\,600\,000 samples. The final sample size depends on the number of light curves to be analyzed. 

We visually inspect all TLCs and select a general baseline model of a quadratic polynomial in time to correct for time-dependent modulations. We also assume that some of the EulerCam observations must have suffered from coordinate drifts of the telescope, hence, we test the application of an additional quadratic polynomial in the telescope drift. The TRAPPIST light curves seemed to have experienced difficulties with the autofocus, as discussed in \citet{NVM2016}, resulting in significant variations of the full-width-half-maximum values for each image. Therefore, we also consider multiplying our light curve models with second-order polynomials with respect to the FWHM values. However, we only choose more complex models over our general (minimal) baseline model, if the derived Bayes factor \citep[e.g.,][]{Schwarz} implies a higher probability. The final baseline model for each light curve is presented in Table \ref{tab:obsinfo}.

We find anomalies due to occulted starspots in the transit light curves obtained with the DFOSC instrument (2015/05/13 and 2015/05/17), in the EulerCam (2011/04/02, 2011/05/15, and 2011/01/31) and TRAPPIST (2011/04/02) observations. The EulerCam (2011/04/02) and the TRAPPIST (2011/04/02) measurements observed the same transit event, hence, they show the same spot. To obtain accurate values for the spot location, size and contrast, we also fit these data sets separately, and compare the results to the values which we derive from the simultaneous analysis.

\subsection{Results}\label{sec:Resfour}

\subsubsection{WASP-41b System Parameters}\label{sec:ResW41}

\begin{figure}
        \centering
	\includegraphics[clip, trim=0.0cm 0.4cm 1cm 0.4cm, width=\columnwidth]{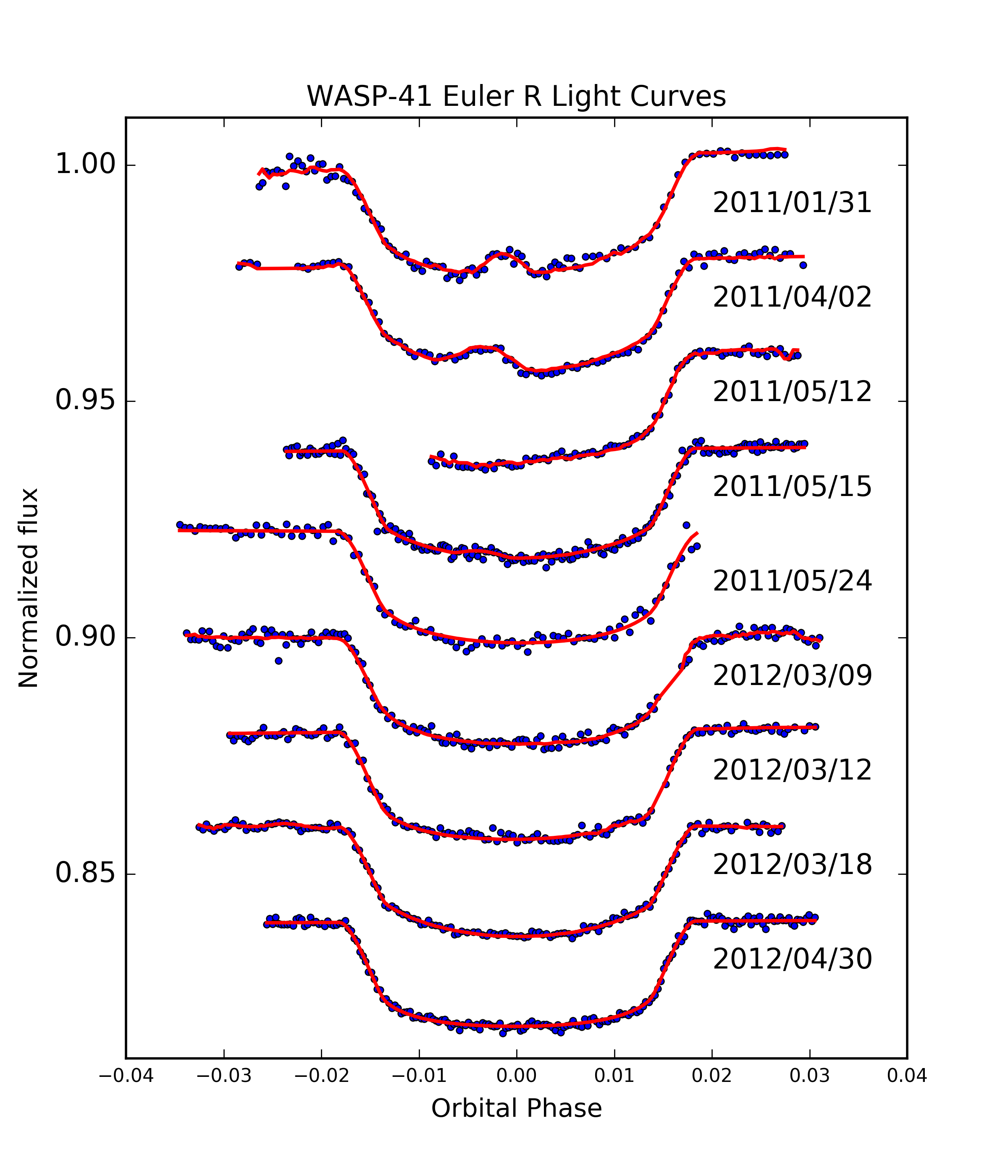}
    \caption{EulerCam light curves (blue dots) and best-fit models (red line) with the corresponding residuals shown in Figure \ref{fig:monires}. The TLCs are presented with their respective date of observation. Note that the light curves with observing dates 2011/01/31, 2011/04/02 and 2011/05/15 were modelled with an occulted starspot feature. Results obtained from the simultaneous analysis are given in Table \ref{tab:simultmulti} and \ref{tab:simultmultiwspots} and the derived spot parameters from the simultaneous and individual analysis are presented in Table \ref{tab:rss} and \ref{tab:resspot}.}
    \label{fig:moni}
\end{figure}

Figures \ref{fig:moni}, \ref{fig:dfosc}, \ref{fig:trappist}, and \ref{fig:RandT82} present the WASP-41b TLCs with their best-fit models and residuals from the individual analyses. Tables \ref{tab:simultmulti} and \ref{tab:simultmultiwspots} then give the results from the simultaneous fitting processes for the two discussed cases (assuming a spot and a spot-free model). We decide to thoroughly present the simultaneous fitting results only, as the results of the individual TLCs agree within one sigma with the results from the combined fit. The only exceptions are the EulerCam TLC (2011/05/12) and the DFOSC R TLC (2013/04/23), which show discrepancies in the transit midtime, planet-to-star radius ratio, relative semi-major axis, and limb darkening coefficients to an extent of 4$\sigma$. This is likely the result of the lack of data points in the first (or second) half of the transit, which directly affects the accurate determination of these transit parameters.

\begin{figure}
        \centering
	\includegraphics[clip, trim=0.0cm 0.4cm 1cm 0.4cm, width=0.85\columnwidth]{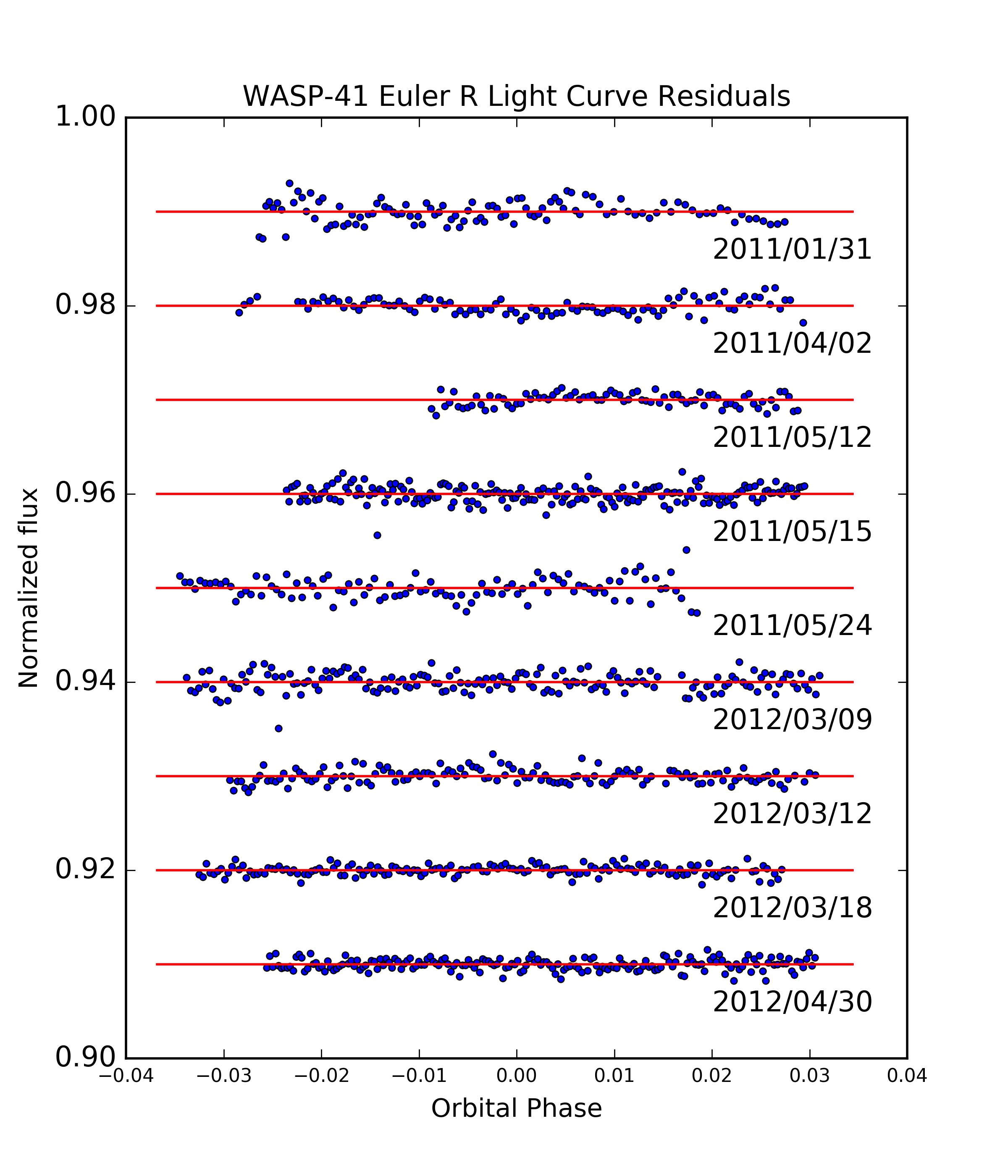}
    \caption{Residuals of the EulerCam light curves. The light curves with observing dates 2011/01/31, 2011/04/02 and 2011/05/15 were modelled with an occulted starspot feature around phase -0.002.}
    \label{fig:monires}
\end{figure}

\begin{figure}
        \centering
	\includegraphics[clip, trim=0.0cm 0.5cm 0.0cm 0.9cm, width=0.85\columnwidth]{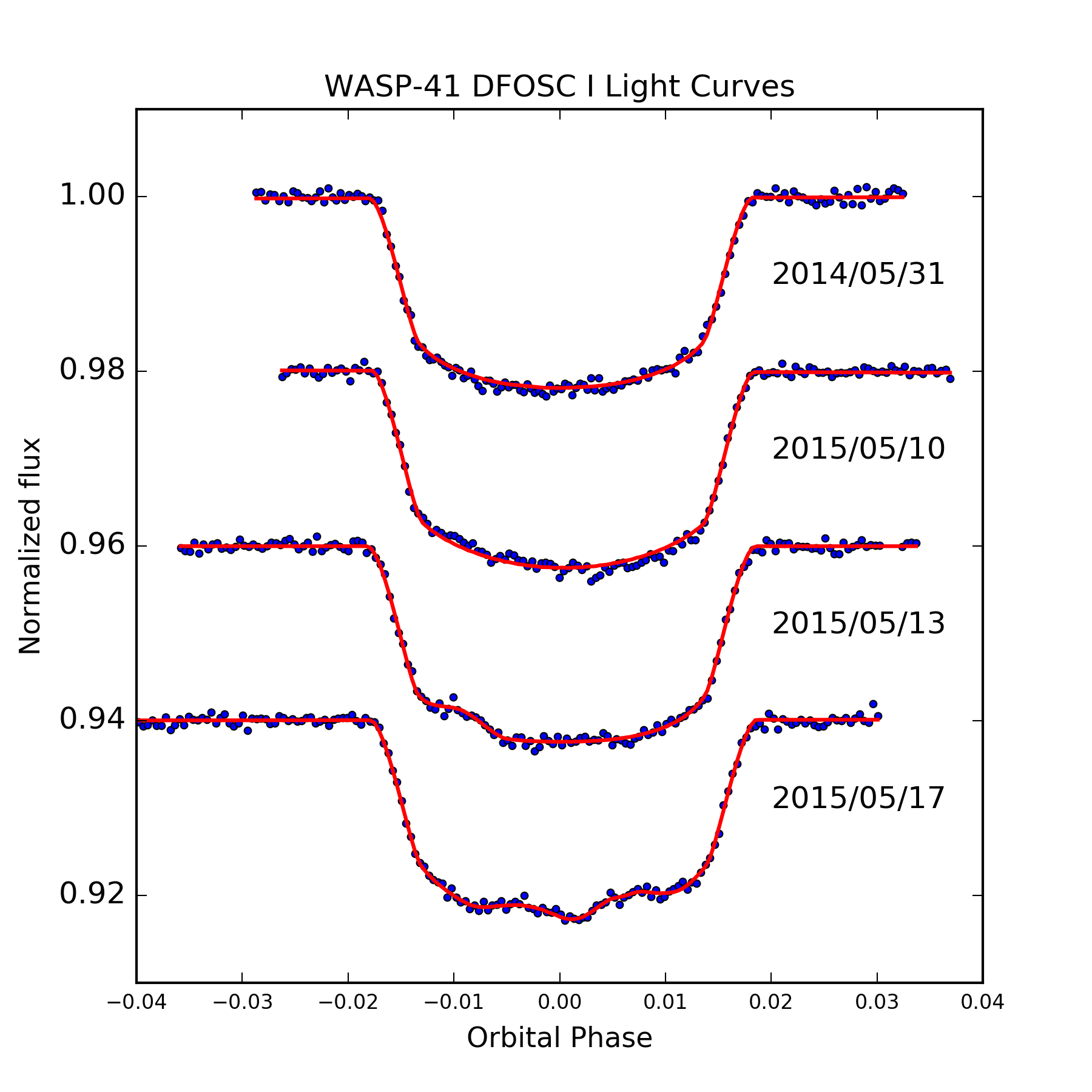}
    \caption{DFOSC I light curves (blue dots) and best-fit models (red line). The corresponding residuals are shown in Figure \ref{fig:dfoscres}. The TLCs are listed with their respective date of observation, and the data with observing date 2015/05/13 and 2015/05/17 show one and two occulted spots, respectively. Results obtained from the simultaneous analysis are given in Table \ref{tab:simultmulti} and \ref{tab:simultmultiwspots} and the derived spot parameters from the simultaneous and individual analysis are presented in Table \ref{tab:rss} and \ref{tab:resspot}.}
    \label{fig:dfosc}
\end{figure}

\begin{figure}
        \centering
	\includegraphics[clip, trim=0.0cm 0.5cm 0.0cm 0.7cm, width=0.85\columnwidth]{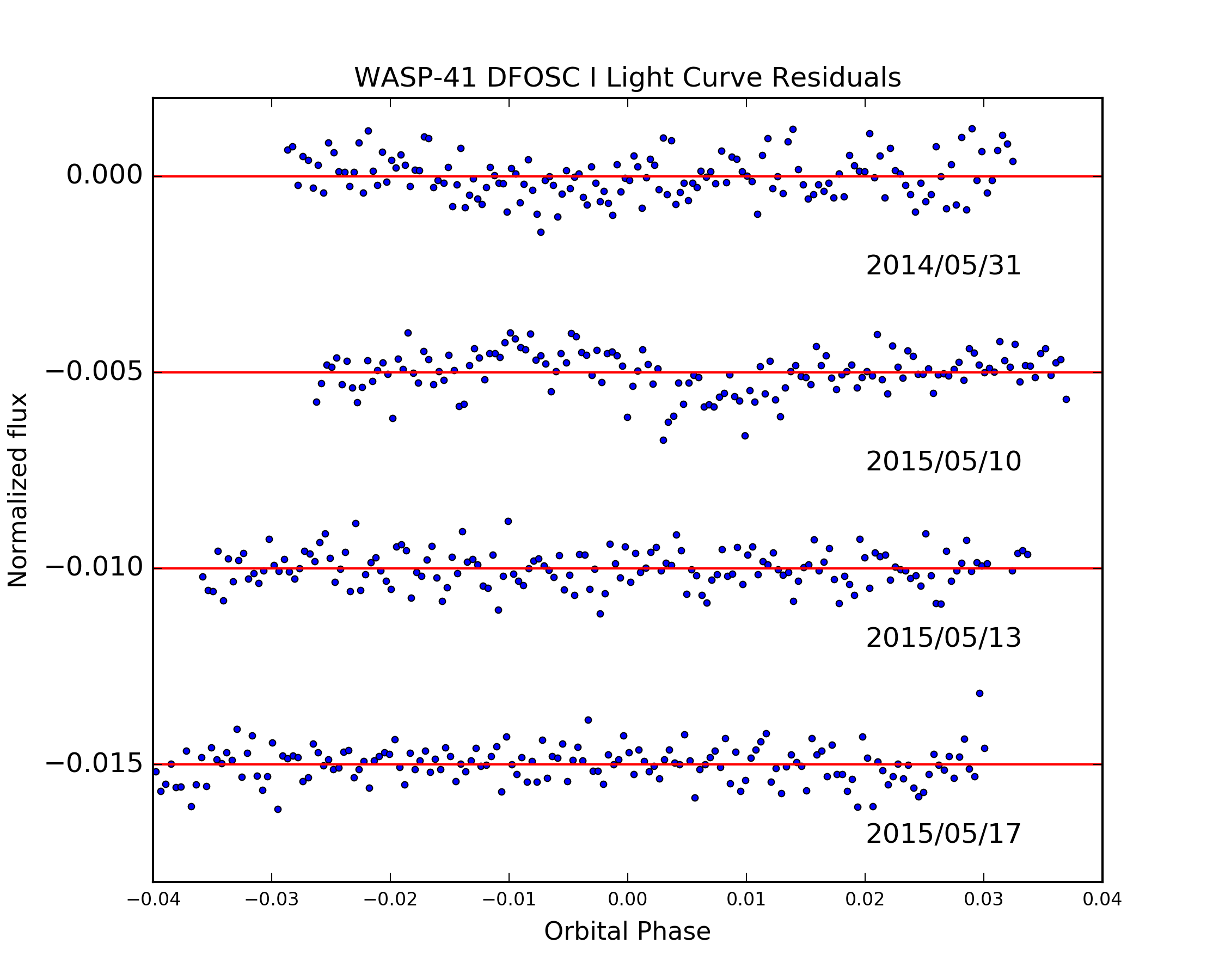}
    \caption{Residuals of the DFOSC I light curves, listed with their respective date of observation. The transit light curves with observing date 2015/05/13 and 2015/05/17 show one (phase -0.01) and two occulted spots (phases -0.005 and 0.005), respectively.}
    \label{fig:dfoscres}
\end{figure}

\begin{figure}
        \centering
	\includegraphics[clip, trim=0.0cm 0.5cm 0.0cm 0.7cm, width=0.85\columnwidth]{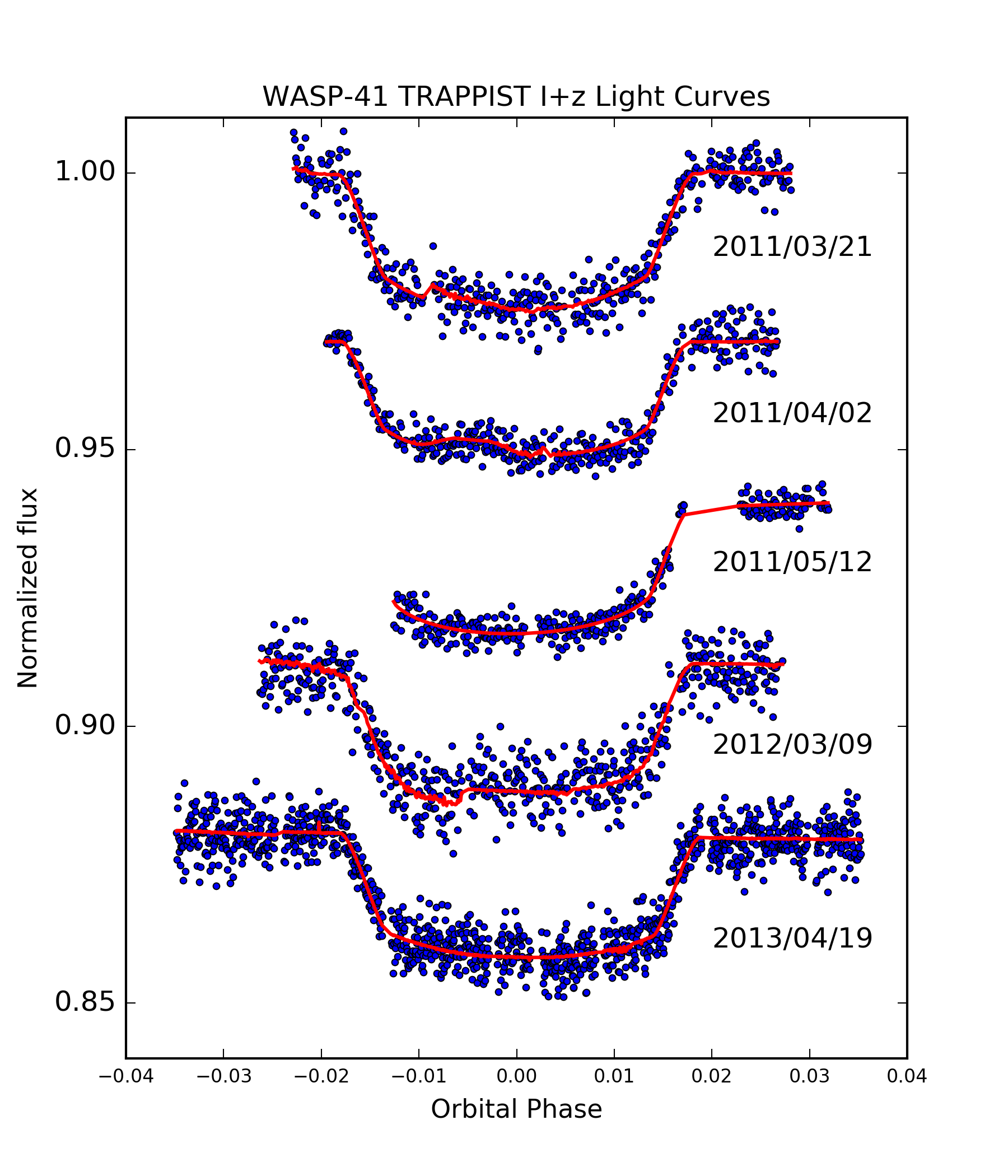}
    \caption{TRAPPIST light curves (blue dots) with the derived best-fit models (red line). The corresponding residuals are shown in Figure \ref{fig:trappistres}. The TLCs are presented with their corresponding observing date. The TLC with the observing date 2011/04/02 shows a starspot at phase -0.005. Results, which we obtained from the simultaneous analysis, are given in Table \ref{tab:simultmulti} and \ref{tab:simultmultiwspots} and the derived spot parameters from the simultaneous and individual analysis are presented in Table \ref{tab:rss} and \ref{tab:resspot}.}
    \label{fig:trappist}
\end{figure}

\begin{figure}
        \centering
	\includegraphics[clip, trim=0.0cm 0.5cm 0.0cm 0.9cm, width=0.85\columnwidth]{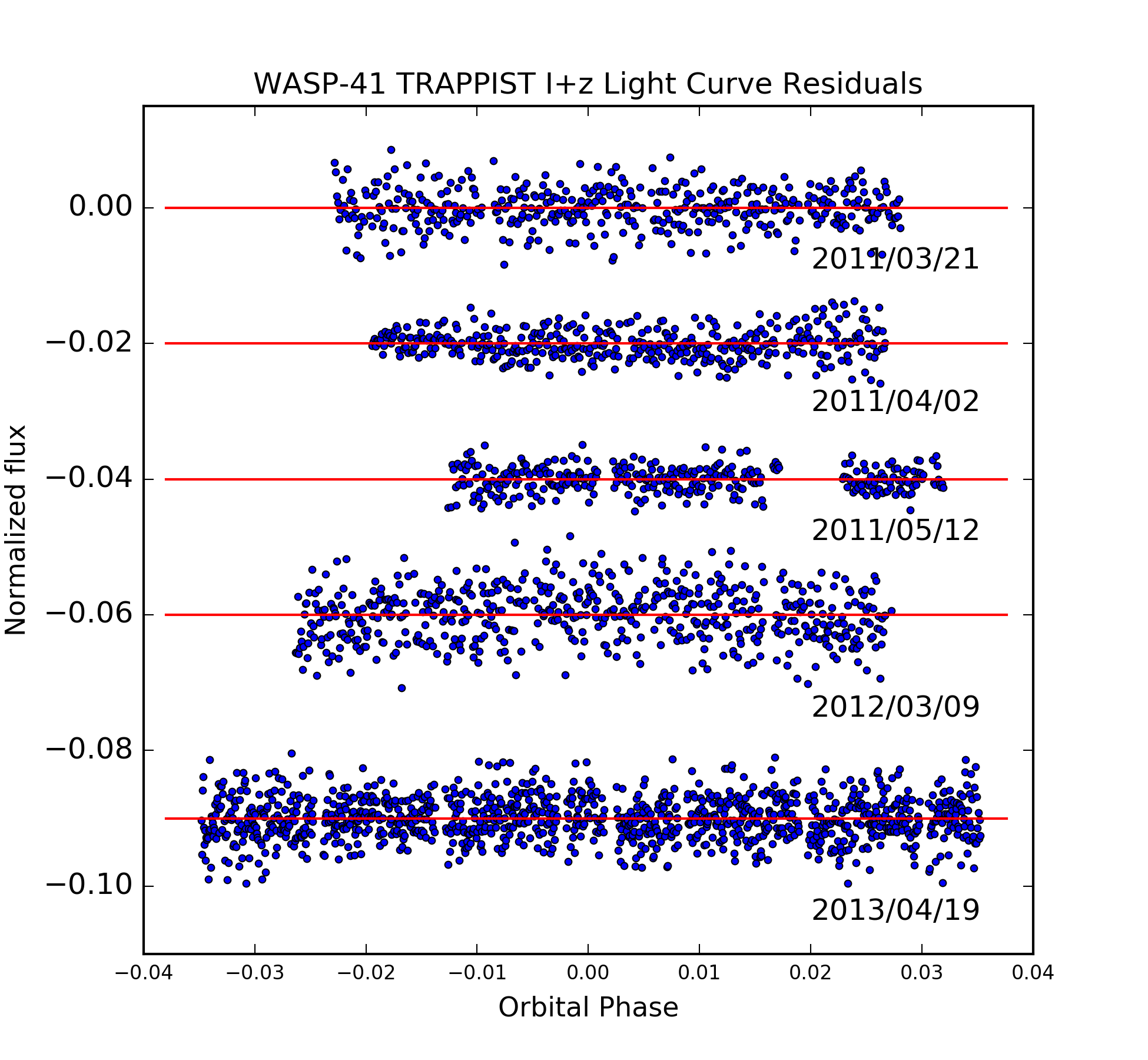}
    \caption{Residuals of the TRAPPIST light curves (blue dots), given with their corresponding observing date. The TLC with the observing date 2011/04/02 shows a starspot at phase -0.005.}
    \label{fig:trappistres}
\end{figure}

\begin{figure}
        \centering
	\includegraphics[clip, trim=0.0cm 0.0cm 0.0cm 0.5cm, width=0.85\columnwidth]{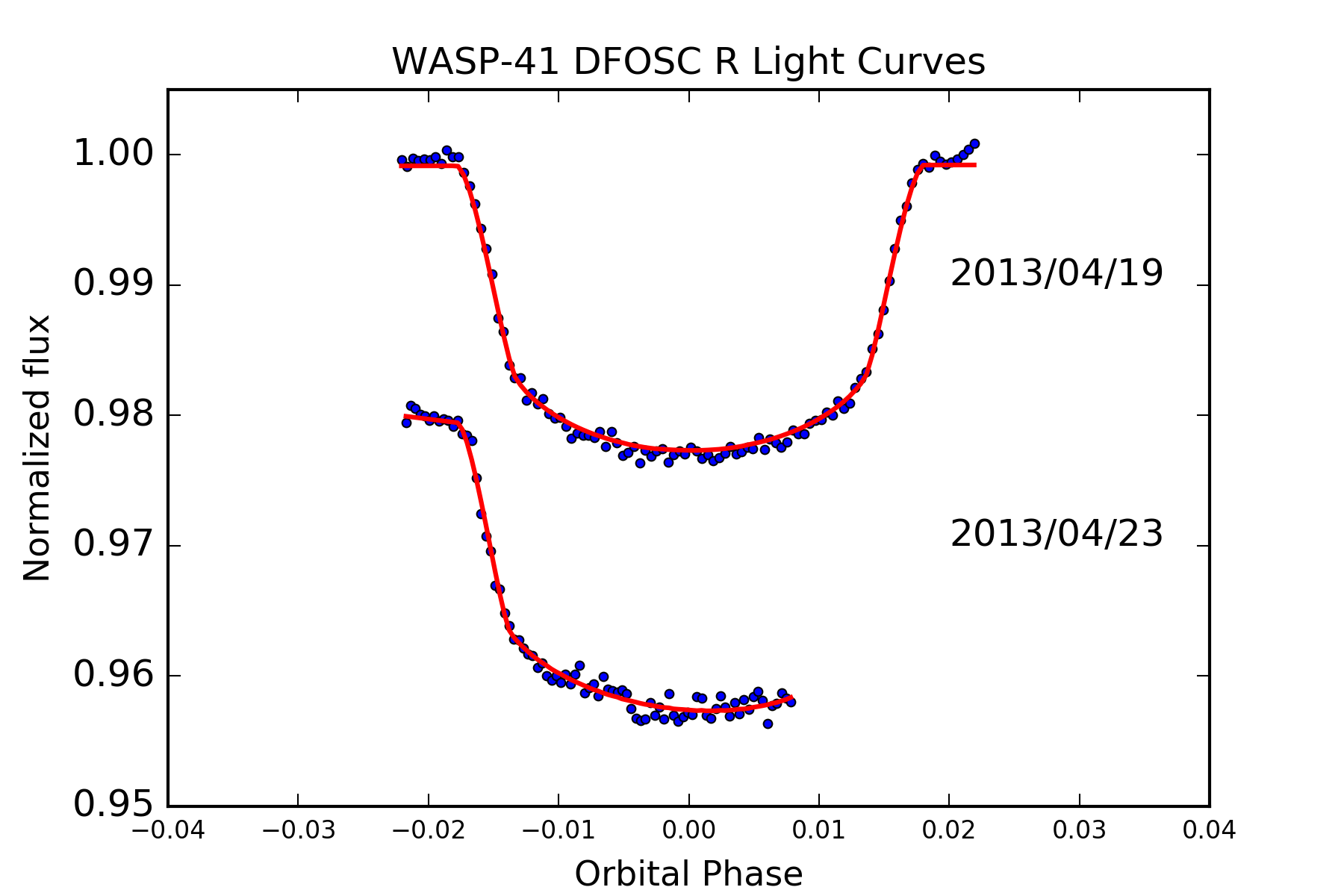}
    \caption{DFOSC R light curves (blue dots) and obtained best-fit models (red line) with the corresponding residuals shown in Figure \ref{fig:RandT82res}. On the right hand side, we indicated the corresponding observing dates. Note that only one of the light curves shows a full transit event. Results obtained from the simultaneous analysis are given in Table \ref{tab:simultmulti} and \ref{tab:simultmultiwspots}.}
    \label{fig:RandT82}
\end{figure}

\begin{figure}
        \centering
	\includegraphics[clip, trim=0.0cm 0.0cm 0.0cm 0.5cm, width=0.85\columnwidth]{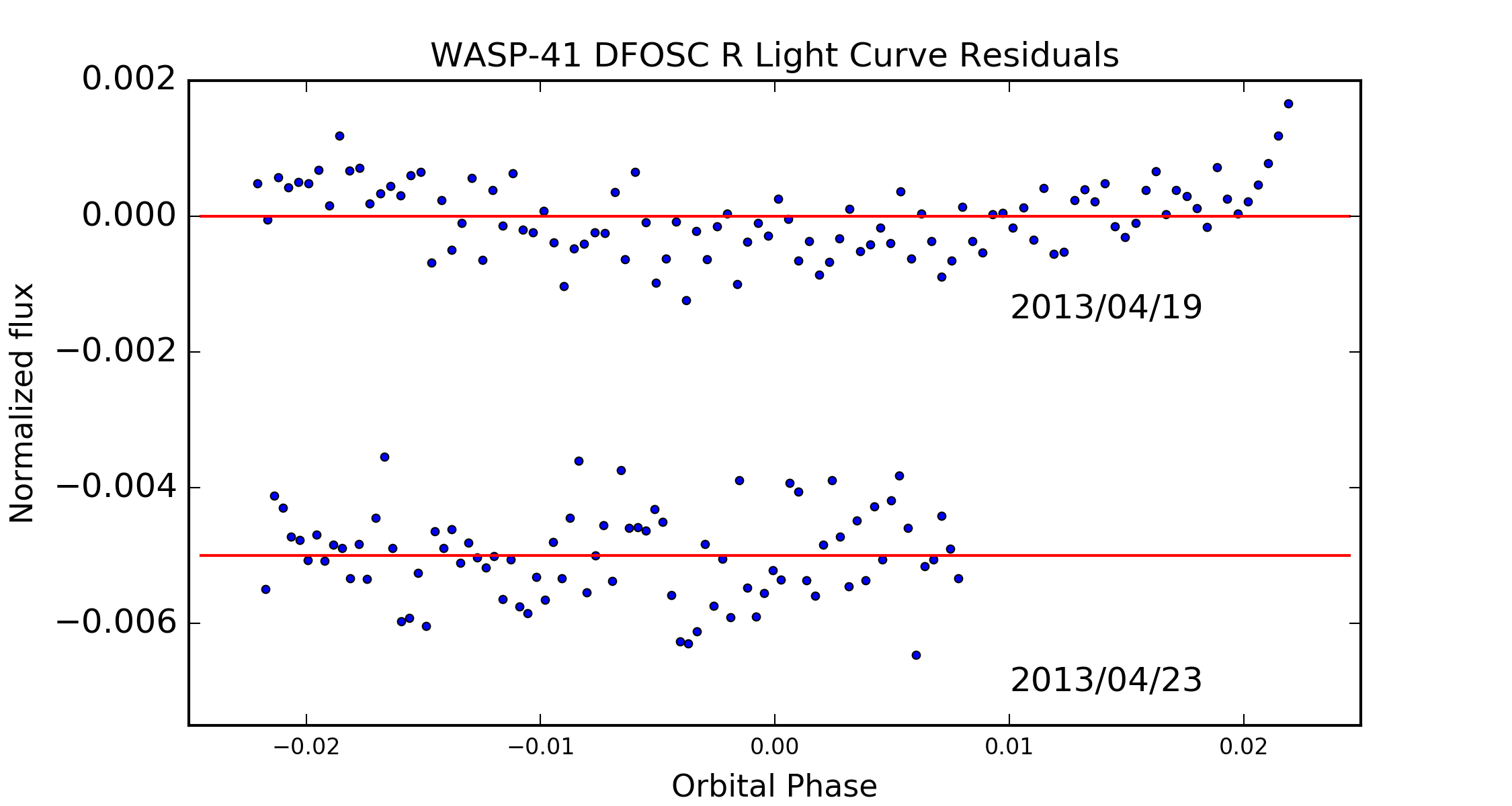}
    \caption{Residuals of the DFOSC R light curves (blue dots) with the corresponding observing dates indicated on the right hand side.}
    \label{fig:RandT82res}
\end{figure}

We compare all our results, from the individual as well as the simultaneous analyses, with those presented in \citet{SW2016} and find that they agree within one sigma (with the exception of the EulerCam TLC (2011/05/12) and DFOSC R (2013/04/23)). We remark also that in the course of the simultaneous analysis, we identify that one of the TLCs observed with TRAPPIST (2011/03/21) leads to an excess value of the transit depth of $r_{\rm{P}}/r_{\rm{S}}\sim0.144$, which differs by almost 2$\sigma$ from the $r_{\rm{P}}/r_{\rm{S}}$ results of the remaining TRAPPIST TLCs. We believe that this must be a result of the TRAPPIST autofocus issues \citep{NVM2016}, as this TLC is affected the most by variations of the FWHM. Therefore, we exclude this light curve from the analysis.

\subsubsection{Spot Modeling Results}\label{sec:ResW41Spots}

At first, we verify if the results derived from the individual and simultaneous analyses are consistent. We find that all obtained spot locations, sizes and contrasts agree within one sigma, which shows that our simultaneous fitting process can also produce reliable spot parameters. We also investigate the EulerCam light curves for possible reappearing spot features. We do not detect reappearing starspots as the observations were taken too far apart when compared to the estimated stellar rotation period of $18.6 \pm 1.5$ days \citep{SW2016} and from rotational modulation, $18.41 \pm 0.05$ days \citep{Maxted2011}.

When we compare our individually and simultaneously derived spot parameters of the DFOSC I light curves (Table \ref{tab:rss} and \ref{tab:resspot}) with the results published by \citet{SW2016}, we find that they agree well within the uncertainties. We also conclude that the spot longitudes are in general well determined, while we find large uncertainties for the spot latitudes. \citet{SW2016} mention that they encountered the same issue and further argue that this is to be expected for systems in which the transit cord is close to the center of the stellar sphere (as it is the case for WASP-41b).

One of the TRAPPIST TLCs (2011/04/02) shows the same spot feature as the EulerCam observation obtained during the same night. We remark that these observations were performed in different filters and furthermore, the TRAPPIST measurements suffered from problems affecting the telescope autofocus, which resulted in large systematics. Yet, the obtained spot parameters (spot longitude, co-latitude, contrast and size) agree within $1.5\sigma$. We also use the derived spot contrasts from the simultaneous multiband measurements of the same starspot (Table \ref{tab:rss}) together with $T_{\rm{eff}}$ from Table \ref{tab:wasp41} to calculate the spot temperatures using equation (1) of \citet{Silva2003}, assuming a black body approximation. We find that the obtained spot temperatures $T_{\rm{spot, TRAPPIST}} = {5296}^{+119}_{-245}$ K and $T_{\rm{spot, EulerCam}} =  {5220}^{+58}_{-126}$ K agree well within their uncertainties. The indicated temperature difference (between the stellar photosphere and the spot) of about 300 K is consistent with literature values obtained for other main-sequence stars, as illustrated in Figure (8) of \citet{Mancini2016}.

\subsubsection{WASP-41b Transmission Spectrum}\label{sec:ResW41}

\begin{figure}
        \centering
	\includegraphics[width=\columnwidth]{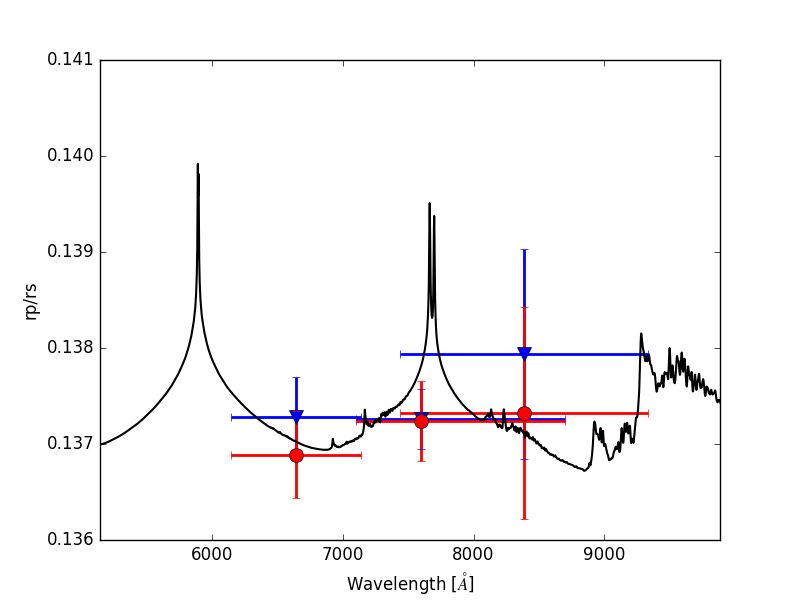}
    \caption{Transmission spectrum obtained from the simultaneous light curve analyses for two cases: The blue triangles show the radius ratios with their respective uncertainties from the analysis of all light curves assuming a spot-free model. The red circles present the radius ratios and unertainties obtained through the analysis of all TLCs, taking starspots into account. We remark that for both scenarios, we excluded the TRAPPIST (2011/03/21) light curve from the analysis.}
    \label{fig:transms}
\end{figure}

In Figure \ref{fig:transms} we show the wavelength dependent planet-to-star radius ratios, which we derive from the simultaneous analyses assuming a spot and a spot-free model. Our results are compatible with a flat transmission spectrum. We compare our data to a model transmission spectrum computed with the \texttt{Pyrat Bay} package (Python Radiative-transfer in a Bayesian framework, \citet[][in prep.]{Patocode}), which is based on \citet{Blecic2016phdThesis} and \citet{Cubillos2016phdThesis}. The model assumes a solar-abundance atmosphere in thermo-chemical equilibrium \citep{BlecicEtal2016apsjTEA}, hydrostatic equilibrium, and an isothermal temperature profile (at the WASP-41b equilibrium temperature), for the system parameters given in Table \ref{tab:wasp41}. The radiative-transfer calculation considers opacities from Na and K \citep{BurrowsEtal2000apjBDspectra}, H$_{2}$O and CO$_{2}$ \citep{RothmanEtal2010jqsrtHITEMP}, collision-induced absorption from H$_{2}$-H$_{2}$ \citep{BorysowEtal2001jqsrtH2H2highT, Borysow2002jqsrtH2H2lowT} and H$_{2}$-He \citep{BorysowEtal1988apjH2HeRT,
BorysowEtal1989apjH2HeRVRT, BorysowFrommhold1989apjH2HeOvertones}, and H$_{2}$ Rayleigh scattering \citep{LecavelierDesEtangsEtal2008aaRayleighHD189}.
While a flat transmission spectrum can be interpreted as a sign of aerosols (clouds or hazes) in the planet's atmosphere, our data do not have the necessary resolution to distinguish a cloudy from a cloud-free atmosphere.

\subsubsection{Impact of Starspots on Transit Parameters}\label{sec:ResW41Spots}

To investigate the impact of neglecting starspots in the analysis, we also model the spotted TLCs assuming they are spot-free. We further use these results to verify (by means of BIC comparison) that models, which take into account starspots, have a higher probability than spot-free models. We then compare the spot and spot-free model results and find that the transit parameters, derived from the simultaneous and for most cases of the individual analyses, agree well within 1.4$\sigma$. The transit depth values derived in the Gunn r' (including Bessell R), the Bessell I and the I+z band, assuming a spot-free model, agree well with those inferred when fitting for starspots. Yet, we find the following discrepancies for the individually analyzed EulerCam TLCs (2011/01/31, 2011/04/02) and the DFOSC I observation (2015/05/13):
\begin{itemize}
\item The results from the spot-free analysis of the EulerCam TLC (2011/04/02) show differences in phaseoffset, $a/R_{\rm{S}}$ and limb darkening coefficients by $<$3$\sigma$, $<$2$\sigma$, and $<$5$\sigma$, respectively. We find similar discrepancies for the DFOSC I (2015/05/13) TLC affecting also the $a/R_{\rm{S}}$ value and the limb darkening coefficients.
\item The EulerCam (2011/01/31) TLC only shows differences in the limb darkening coefficients to an extent of $<$2$\sigma$.
\end{itemize}
The impact of stellar spots on the determination of limb darkening coefficients (LDCs), the relative semi-major axis $a/R_{\rm{S}}$, and the transit timing and duration is a known issue, and has been investigated by several authors \citep[e.g.,][]{Silva2010, Ballerini2012,TR2013, Osh2013, Csiz2013}. Especially starspots, which are located at the limb of the star, can bias the transit ingress (or egress) time of the planet. This directly affects the measurement of the total transit duration and hence, the relative semi-major axis $a/R_{\rm{S}}$. It is thus important to account for spot features in TLC fitting processes to derive precise transit parameters, and when studying transit timing variations and transmission spectra.

\section{Summary and Conclusions}\label{sec:five}

We have presented \texttt{PyTranSpot}, a routine designed to model transit light curves with stellar activity features. \texttt{PyTranSpot} uses a pixellation approach to model the transiting planet, stellar limb darkening and starspots on the stellar surface. We have merged \texttt{PyTranSpot} with the MCMC framework developed by \citet[][]{Lendlcode} to derive accurate exoplanet transmission spectra in the presence of starspots and correlated instrumental noise.

We validated our routine by analyzing eleven synthetic light curves of four different star-planet configurations, including 6 synthetic light curves which show anomalies due to occulted starspots. By comparing our derived results with the original system parameters, we found that \texttt{PyTranSpot} could reproduce the properties of our synthetic systems.

We further performed a multi-wavelength analysis of 20 transit light curves of the well-studied system WASP-41b using archival and yet unpublished data. From this data set, 7 TLCs were affected by starspot occultations. We analyzed the light curves simultaneously for two cases (analysis of all TLC assuming a spot and a spot-free model) as well as individually. We found that our derived results agree well within one sigma with the values given in the literature. In our study, assuming a spot-free model for the spotted TLCs did not seem to have a significant impact on the determination of the transit depth. However, we identified that not taking into account stellar spots in the (individual) TLC analyses affected measurements of the limb darkening coefficients, the relative semi-major axis $a/R_{\rm{S}}$ and the transit midtime. These results confirm findings and predictions from various authors \citep[e.g.,][]{Silva2010, SO2011, Ballerini2012, TR2013, Osh2013, Csiz2013}. For each simultaneous analysis, we additionally obtained three wavelength dependent $r_{\rm{P}}/r_{\rm{S}}$ values for WASP-41b covering a range of about 6200-9200 $\AA$. We do not observe any significant variation of the transit depth with wavelength, however, our data are fully compatible both with a cloud-free and cloudy (i.e., flat) transmission spectrum. From the simultaneous multiband observation of the same starspot in the TLCs EulerCam and TRAPPIST (2011/04/02), we could further derive a temperature difference between the stellar photosphere and the starspot of about 300K, which is consistent with literature values for similar dwarf stars \citep[e.g.,][]{Mancini2016}.

We conclude that the outcome of our light curve analyses illustrates the importance of accounting for stellar activity features in TLCs for the correct interpretation of exoplanet transit parameters, transit timing variations and transmission spectra. Furthermore, having simultaneous multiband observations of occulted starspots can help constrain a starspot's temperature, disentangling the correlation between spot size and temperature.

\begin{acknowledgements}

We acknowledge the Austrian Forschungsf\"orderungsgesellschaft FFG projects ``RASEN'' P847963. We also thank the anonymous referee for helping to improve the original manuscript. We thank contributors to Numpy \citep{Nump}, SciPy \citep{Sci}, Matplotlib \citep{Matlib2007}, the Python Programming Language, and contributors to the free and open-source community. I.J. thanks M. Oshagh for fruitful discussions and J. Southworth for providing the corrected spot parameter. H. Lammer and M. Lendl also acknowledge support from the FWF project P25256-N27 ``Characterizing Stellar and Exoplanetary Environments via Modelling of Lyman-$\alpha$ Transit Observations of Hot Jupiters".

\end{acknowledgements}

\bibliography{example}

\begin{appendix}

\section{Synthetic light curve residuals}\label{sec:resplots}

\begin{figure}[!h]
        \centering
	\includegraphics[width=0.85\columnwidth]{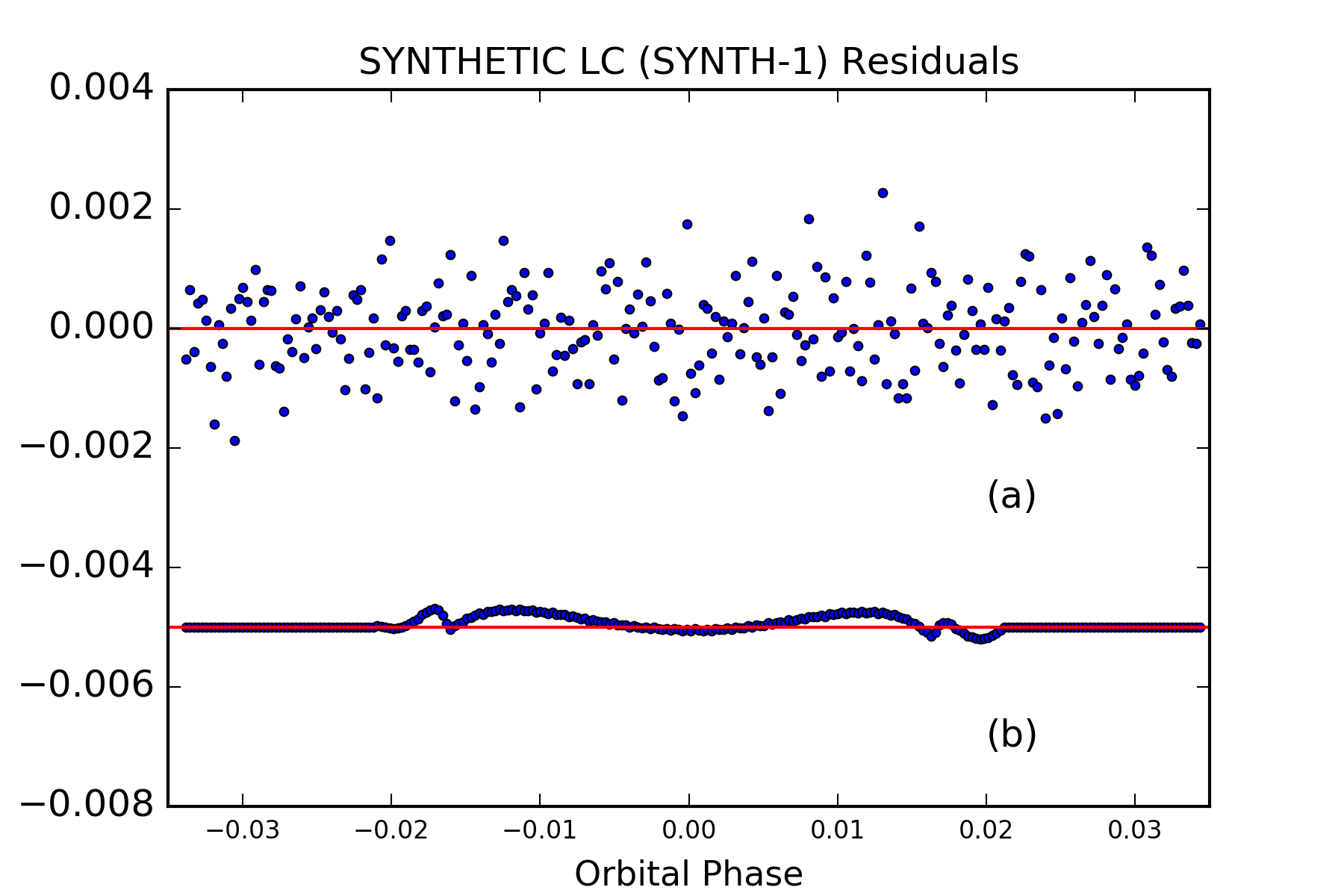}
    \caption{(O-C) residuals (a) and the difference between original and best-fit light curve model (b) of SYNTH-1. The obtained photometric parameters can be found in Table \ref{tab:ressynth}.}
    \label{fig:Synth1res}
\end{figure}

\begin{figure}[!h]
        \centering
	\includegraphics[width=0.85\columnwidth]{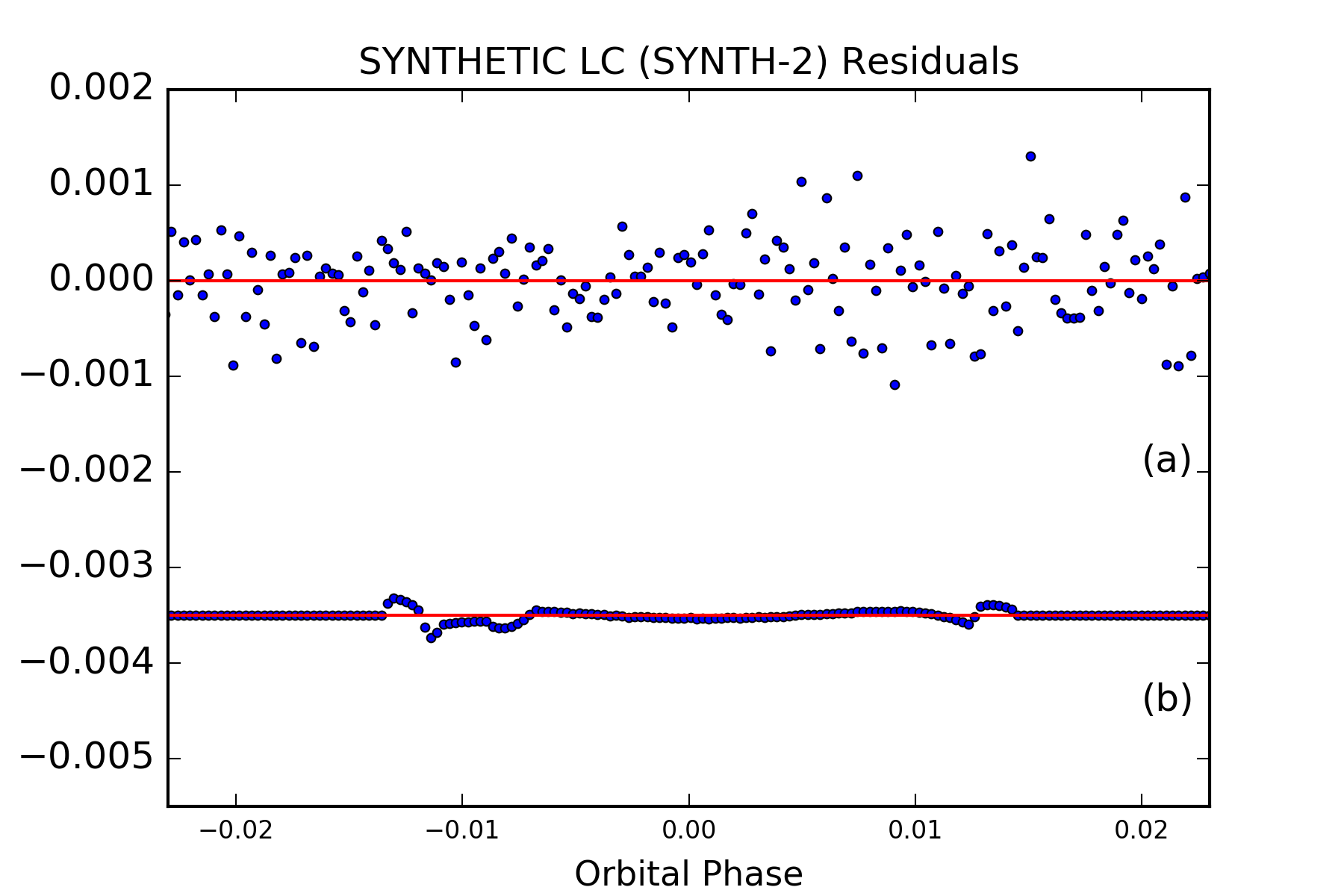}
    \caption{(O-C) residuals (a) and the difference between original and best-fit light curve model (b) of system SYNTH-2. The SYNTH-2 light curve shows an occulted starspot at the limb of the star around phase -0.01. The obtained photometric and spot parameters can be found in Tables \ref{tab:ressynth} and \ref{tab:spotsyn}.}
    \label{fig:Synth2res}
\end{figure}

\begin{figure}[!h]
        \centering
	\includegraphics[clip, trim=0.0cm 0.7cm 0.0cm 0.7cm, width=0.85\columnwidth]{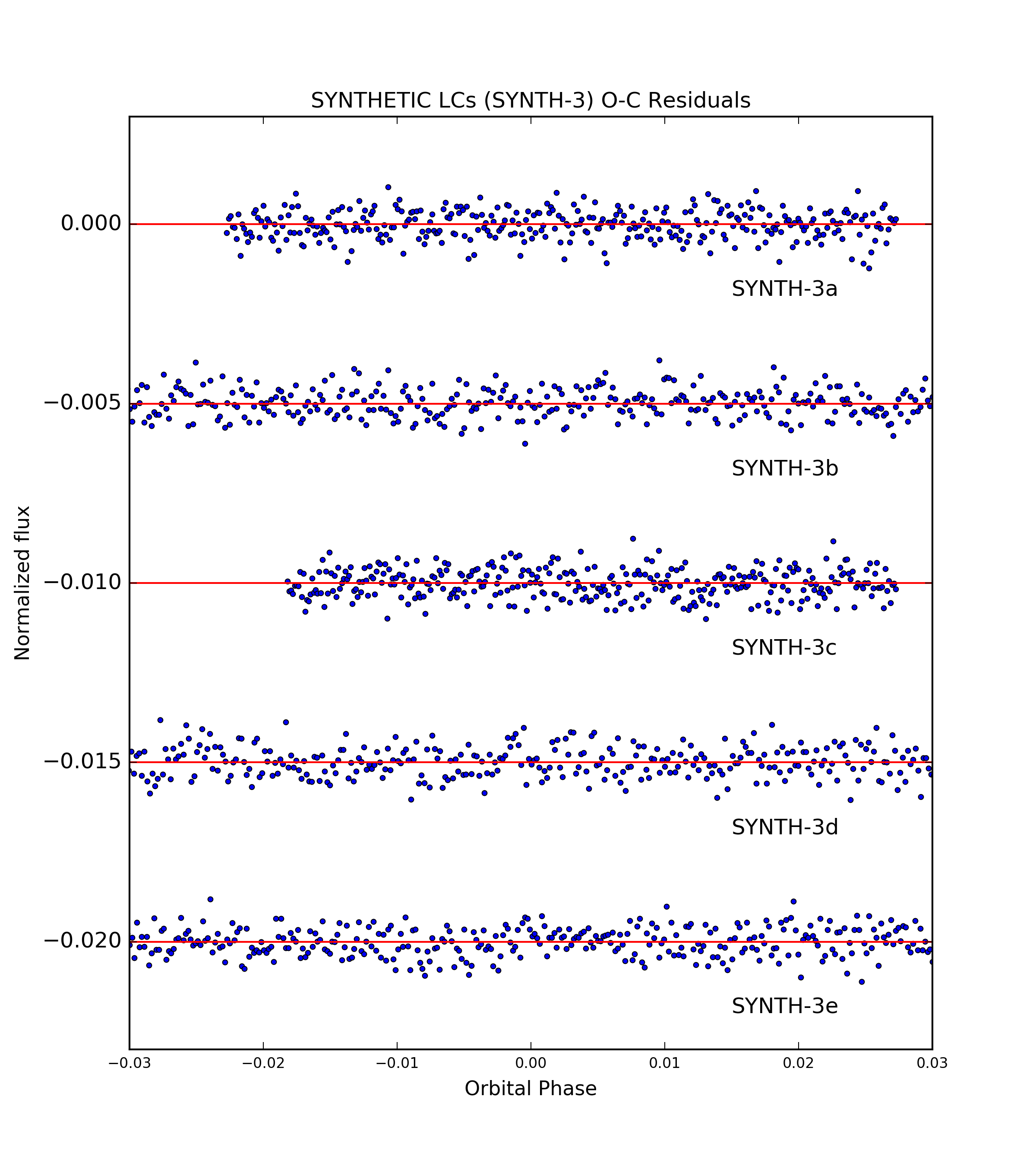}
    \caption{The (O-C) residuals of the synthetic light curves SYNTH-3a to SYNTH-3e. SYNTH-3e shows a starspot anomaly located close to the limb of the star around phase -0.005. The obtained photometric and spot parameters can be found in Tables \ref{tab:ressynth} and \ref{tab:spotsyn}.}
    \label{fig:Synth3res}
\end{figure}

\begin{figure}[!h]
        \centering
	\includegraphics[clip, trim=0.0cm 0.7cm 0.0cm 0.7cm, width=0.85\columnwidth]{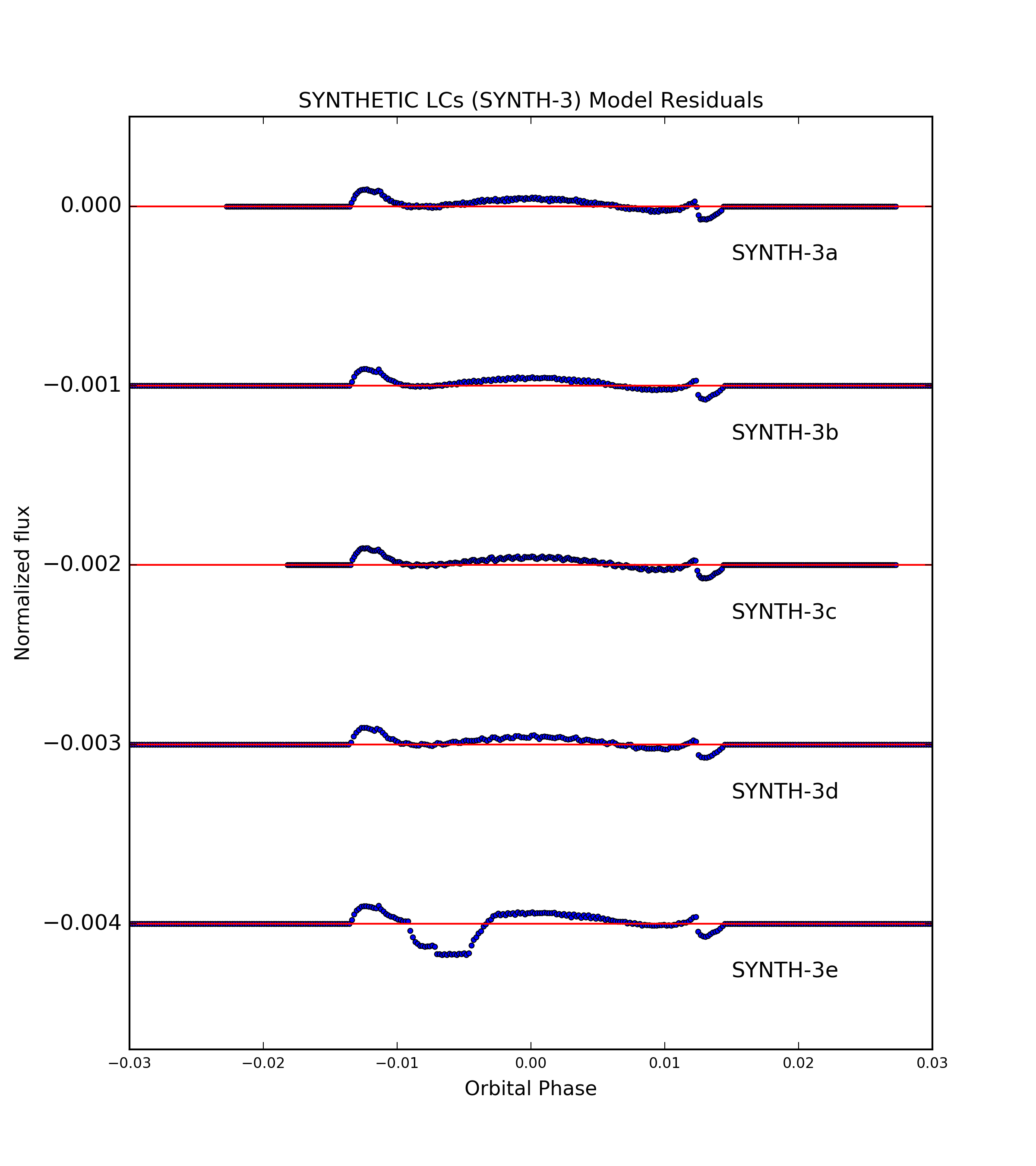}
    \caption{Differences between the original and best-fit light curve models of SYNTH-3a, -3b, -3c, and -3e.}
    \label{fig:Synth3mod}
\end{figure}

\begin{figure}[!h]
        \centering
	\includegraphics[clip, trim=0.0cm 0.7cm 0.0cm 0.7cm, width=0.85\columnwidth]{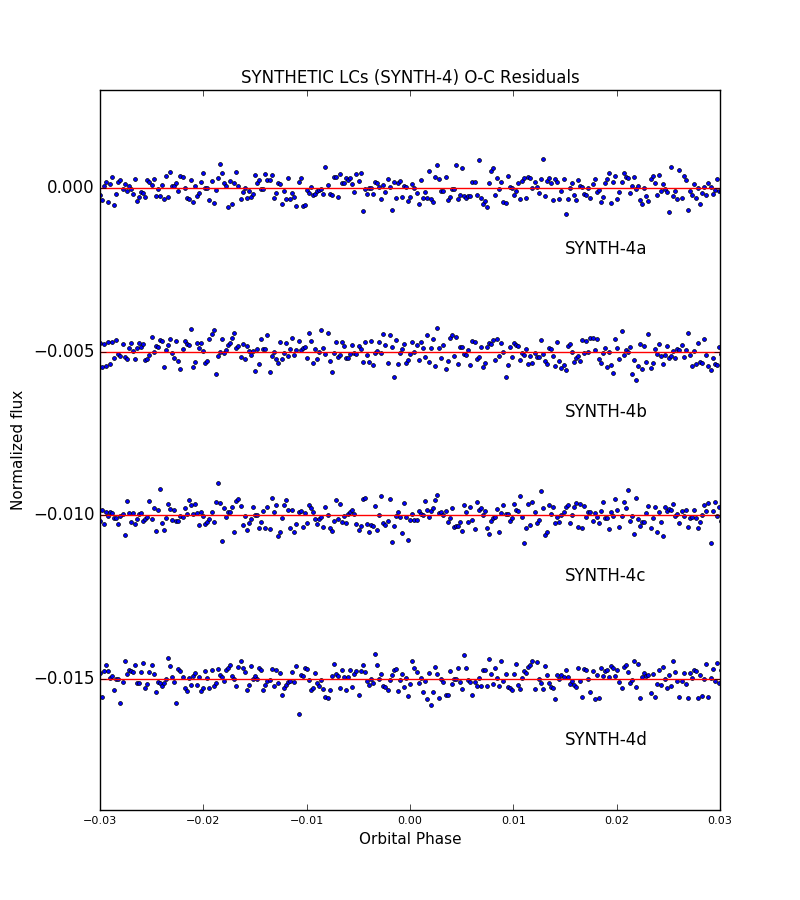}
    \caption{(O-C) residuals of the light curves SYNTH-4a to SYNTH-4d. All TLCs show the same transit event and occulted starspot around phase -0.002. The obtained photometric and spot parameters can be found in Tables \ref{tab:ressynth4} and \ref{tab:spotsyn4}.}
    \label{fig:Synth4res}
\end{figure}

\begin{figure}[!h]
        \centering
	\includegraphics[clip, trim=0.0cm 0.7cm 0.0cm 0.7cm, width=0.85\columnwidth]{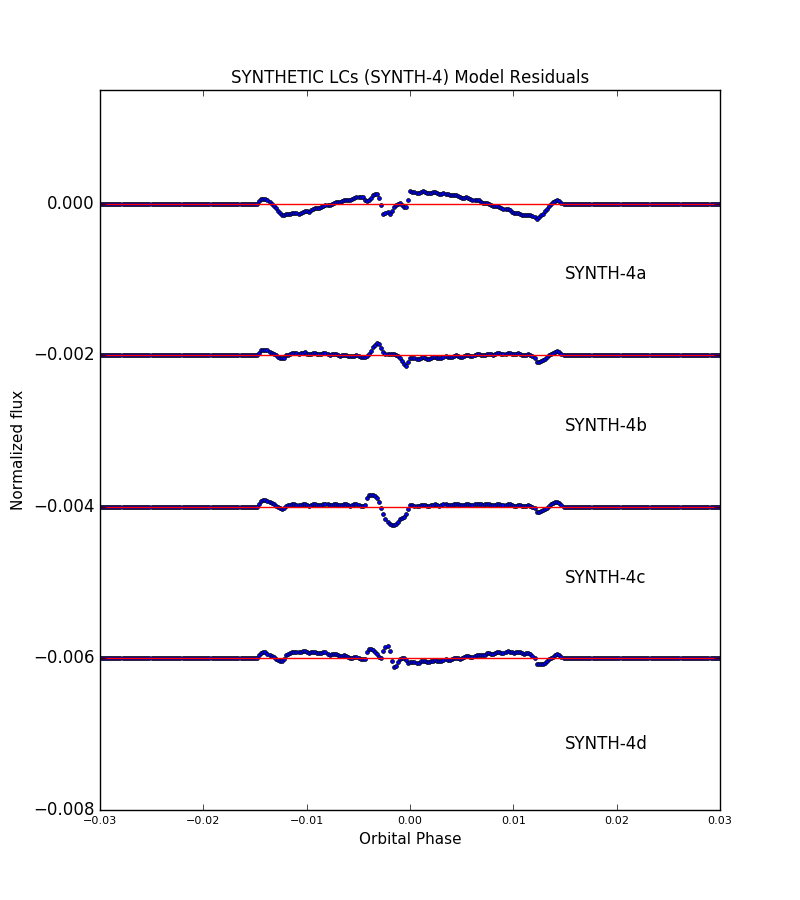}
    \caption{Differences between the original and best-fit light curve models of SYNTH-4a, -4b, -4c, and -4d.}
    \label{fig:Synth4mod}
\end{figure}

\cleardoublepage

\section{Derived Transit and Spot Parameters}\label{sec:params}
\renewcommand{\arraystretch}{1.2}
\begin{table*}[!h]
  \centering 
   \small
\caption{Derived best fit values resulting from the individual (SYNTH-1a and SYNTH-2a) and simultaneous (SYNTH-3a to SYNTH-3e) analysis. Results for the spot parameters can be found in Table \ref{tab:spotsyn}. The parameters are given with their respective one sigma uncertainties.}\label{tab:ressynth}
	\begin{tabular}{lcccccc} 
		\hline
                  Light Curve & phase & $\frac{r_{\rm{P}}}{r_{\rm{S}}}$ & $i$ (\degr) & $\frac{a}{R_{\rm{S}}}$ & $u_{\rm{1}}$ & $u_{\rm{2}}$  \\
		   ID & offset &  &  &  & & \\
		\hline
		SYNTH-1a & $ 0.000098^{\scriptstyle{+0.000052}}_{-0.000051} $  & $ 0.1110^{+0.0031}_{-0.0016} $   & $ 88.0466^{+1.2191}_{-1.2529} $  & $ 8.1146^{+0.2430}_{-0.3975}  $ & $ 0.3969^{+0.0922}_{-0.0887} $ & $  0.4726^{+0.1459}_{-0.2371} $ \\ 
		SYNTH-2a & $ 0.000478^{+0.000074}_{-0.000074} $  & $ 0.0689^{+0.0014}_{-0.0011} $   & $ 88.5547^{+0.5992}_{-0.7075} $  & $ 11.6518^{+0.3412}_{-0.5399} $ & $ 0.6251^{+0.1446}_{-0.1411} $ & $ 0.1817^{+0.2255}_{-0.2599}$    \\
		SYNTH-3a to -3e & $ 0.000411^{+0.00002}_{-0.00002} $  & $ 0.0697^{+0.0008}_{-0.0007}$   & $ 88.6074^{+0.5635}_{-0.4244}$  & $ 11.8357^{+0.3057}_{-0.3128} $ & $ 0.6818^{+0.0577}_{-0.0542} $ & $  -0.0305^{+0.1034}_{-0.1055}$    \\
	\hline
	\end{tabular}
\end{table*}

\renewcommand{\arraystretch}{1.2}
\begin{table*}[!h]
  \centering 
   \small
\caption{Derived best fit transit parameters resulting from the simultaneous analysis of SYNTH-4a to SYNTH-4d. Results for the spot parameters and filter dependent limb darkening coefficients can be found in Table \ref{tab:spotsyn4}. The parameters are given with their respective one sigma uncertainties.}\label{tab:ressynth4}
	\begin{tabular}{lcccc} 
		\hline
                  Light Curve & phase & $\frac{r_{\rm{P}}}{r_{\rm{S}}}$ & $i$ (\degr) & $\frac{a}{R_{\rm{S}}}$ \\
		   ID & offset &  &  &  \\
		\hline
		SYNTH-4a to 4d & $ -0.000014^{+0.000015}_{-0.000014} $  & $ 0.0921^{+0.0006}_{-0.0005}$   & $ 88.79^{+0.3264}_{-0.2897}$  & $ 11.51^{+0.1487}_{-0.1567} $ \\
	\hline
	\end{tabular}
\end{table*}

{
\renewcommand{\arraystretch}{1.2}
\begin{table*}[!h]
  \centering
  \small
\caption{Derived spot parameters for the synthetic light curves SYNTH-2a, SYNTH-3e. Parameters are given with their respective one sigma uncertainties.}
	\begin{tabular}{lccccc} 
		\hline
                  Light Curve & Spot No. & Longitude & Co-Latitude & Size & Contrast \\
		 ID &  & $\theta$  (\degr)  & $\phi$ (\degr)  & $\alpha$ (\degr) & $\rho_{\rm{spot}}$ \\
		\hline
		SYNTH-2a  & 1 & $ -55.34^{+2.39}_{-2.10} $ & $ 72.55^{+5.23}_{-7.13} $  & $ 16.08^{+1.37}_{-2.17}$  & $ 0.771^{+0.038}_{-0.039} $  \\
		SYNTH-3e & 1 & $ -30.22^{+0.76}_{-0.78}  $ & $ 73.24^{+8.15}_{-8.33} $  & $ 12.43^{+2.29}_{-1.40}$  & $ 0.736^{+0.024}_{-0.027}$   \\
	\hline
	\end{tabular}
  \label{tab:spotsyn}
\end{table*}}

{
\renewcommand{\arraystretch}{1.2}
\begin{table*}[!h]
  \centering
  \small
\caption{Derived spot parameters for the synthetic light curves SYNTH-4a to SYNTH-4d. Parameters are given with their respective one sigma uncertainties.}
	\begin{tabular}{lccccccc} 
		\hline
                  Light Curve & Spot No. & Longitude & Co-Latitude & Size & Contrast & $u_{\rm{1}}$ & $u_{\rm{2}}$ \\
		 ID &  & $\theta$  (\degr)  & $\phi$ (\degr)  & $\alpha$ (\degr) & $\rho_{\rm{spot}}$  & & \\
		\hline
		SYNTH-4a (B filter)  & 1 & $ -9.67^{+0.32}_{-0.32} $ & $ 77.69^{+3.60}_{-3.63} $  & $ 4.47^{+0.98}_{-0.55}$  & $  0.507^{+0.0994}_{-0.0785} $ & $ 0.6610^{+0.0656}_{-0.0684} $ & $ 0.2110^{+0.1128}_{-0.1053} $   \\
		SYNTH-4b (V filter) & 1 & $ -9.71^{+0.36}_{-0.38}  $ & $ 77.42^{+4.29}_{-3.94} $  & $ 4.05^{+0.92}_{-0.58}$ & $  0.509^{+0.1188}_{-0.0886} $  & $ 0.3921^{+0.0638}_{-0.0691}$ & $ 0.3499^{+0.1264}_{-0.1014} $  \\
		SYNTH-4c (R filter)  & 1 & $ -9.42^{+0.41}_{-0.82} $ & $ 77.25^{+5.36}_{-4.99} $  & $ 3.90^{+1.17}_{-0.65}$   & $  0.473^{+0.1123}_{-0.0549} $ & $ 0.2622^{+0.0664}_{-0.0770} $ & $0.4514^{+0.1340}_{-0.1446} $  \\
		SYNTH-4d (I filter) & 1 & $ -9.85^{+0.50}_{-0.55}  $ & $ 78.46^{+5.52}_{-5.86} $  & $ 5.14^{+1.69}_{-2.15}$  & $  0.514^{+0.1524}_{-0.0944} $  & $ 0.1932^{+0.0795}_{-0.0882}$ & $ 0.4333^{+0.1509}_{-0.1385} $  \\
	\hline
	\end{tabular}
  \label{tab:spotsyn4}
\end{table*}}

{
\renewcommand{\arraystretch}{1.2}
\begin{table*}[!h]
	\centering
	\caption{Derived best fit parameters of the simultaneous multiband WASP-41b TLC analysis, assuming a spot-free model for all TLCs. Parameters are presented with the corresponding one sigma uncertainties.}\label{tab:simultmulti}
	\begin{tabular}{lcc} 
		\hline
		Parameter Description and Unit & Symbol & WASP-41b system \\
		\hline
		Phase offset & & $ 0.000160^{+0.00001}_{-0.00001} $ \\
		Ratio of fractional radii & $r_{\rm{P}}/r_{\rm{S}}$ & $ 0.136$ [fixed]   \\
		Fitted offset from $r_{\rm{P}}/r_{\rm{S}}$ (I filter)  &  & $ 0.001262^{+0.000311}_{-0.000323} $ \\
		Fitted offset from $r_{\rm{P}}/r_{\rm{S}}$ (I+z filter)  &  & $ 0.001939^{+0.001096}_{-0.001102} $ \\
		Fitted offset from $r_{\rm{P}}/r_{\rm{S}}$ (r' filter)  &  & $ 0.001285^{+0.000410}_{-0.000385}$ \\
                  Orbital inclination (\degr) & $i$ & $ 88.7 $ [fixed]  \\
		Semi-major axis in terms of stellar radius & $a/R_{\rm{S}}$ & $ 9.818^{+0.0147}_{-0.0152} $ \\
                  Planetary orbital period (d) & ${P}_{\rm{orb}}$ & $ 3.05 $ [fixed]   \\
                  Linear LDC (I filter) & $u_{\rm{1,i}}$ &  $ 0.4341^{+0.0261}_{-0.0269} $  \\
                  Quadratic LDC (I filter) &  $u_{\rm{2,i}}$ & $ 0.0943^{+0.0513}_{-0.0495}$  \\
                  Linear LDC (I+z filter) & $u_{\rm{1,iz}}$ &  $ 0.3770^{+0.0654}_{-0.0670}$ \\
                  Quadratic LDC (I+z filter) &  $u_{\rm{2,iz}}$ & $ 0.1007^{+0.1121}_{-0.1104} $  \\
                  Linear LDC (r' filter) & $u_{\rm{1,r}}$ &  $0.4360^{+0.0215}_{-0.0217}$  \\
                  Quadratic LDC (r' filter) &  $u_{\rm{2,r}}$ & $ 0.1657^{+0.0427}_{-0.0423}$  \\
                  Reduced Chi Squared &  $\chi_{\rm{red}}^2$ & $ 1.36 $   \\		
		\hline
	\end{tabular}
\end{table*}}

{
\renewcommand{\arraystretch}{1.2}
\begin{table*}[!h]
	\centering
	\caption{Derived best fit parameters of the simultaneous multiband transit and starspot LC analysis of WASP-41b. Parameters are given with the corresponding one sigma uncertainties. The obtained spot parameters are presented in Table \ref{tab:rss}.}\label{tab:simultmultiwspots}
	\begin{tabular}{lcc} 
		\hline
		Parameter Description and Unit & Symbol & WASP-41b system \\
		\hline
		Phase offset & & $ 0.000172^{+0.00001}_{-0.00001} $ \\
		Ratio of fractional radii & $r_{\rm{P}}/r_{\rm{S}}$ & $ 0.136$ [fixed]   \\
		Fitted offset from $r_{\rm{P}}/r_{\rm{S}}$ (I filter)  &  & $ 0.001239^{+0.000416}_{-0.000401} $ \\
		Fitted offset from $r_{\rm{P}}/r_{\rm{S}}$ (I+z filter)  &  & $ 0.001320^{+0.001103}_{-0.001092} $ \\
		Fitted offset from $r_{\rm{P}}/r_{\rm{S}}$ (r' filter)  &  & $ 0.000889^{+0.000451}_{-0.000504}$ \\
                  Orbital inclination (\degr) & $i$ & $ 88.7 $ [fixed]  \\
		Semi-major axis in terms of stellar radius & $a/R_{\rm{S}}$ & $ 9.795^{+0.0194}_{-0.0188} $ \\
                  Planetary orbital period (d) & ${P}_{\rm{orb}}$ & $ 3.05 $ [fixed]   \\
                  Linear LDC (I filter) & $u_{\rm{1,i}}$ &  $ 0.3775^{+0.0356}_{-0.0357} $  \\
                  Quadratic LDC (I filter) &  $u_{\rm{2,i}}$ & $ 0.2247^{+0.0667}_{-0.0680}$  \\
                  Linear LDC (I+z filter) & $u_{\rm{1,iz}}$ &  $ 0.4014^{+0.0684}_{-0.0701}$ \\
                  Quadratic LDC (I+z filter) &  $u_{\rm{2,iz}}$ & $ 0.1165^{+0.1244}_{-0.1134} $  \\
                  Linear LDC (r' filter) & $u_{\rm{1,r}}$ &  $0.4602^{+0.0274}_{-0.0289}$  \\
                  Quadratic LDC (r' filter) &  $u_{\rm{2,r}}$ & $ 0.1743^{+0.0548}_{-0.0555}$  \\
                  Reduced Chi Squared &  $\chi_{\rm{red}}^2$ & $ 1.05 $   \\		
		\hline
	\end{tabular}
\end{table*}}

{
\renewcommand{\arraystretch}{1.4}
\begin{table*}[!h]
  \centering
  \small
\caption{Derived spot parameters for the simultaneously analyzed WASP-41b light curves. Parameters are given with their respective one sigma uncertainties.}\label{tab:rss}
	\begin{tabular}{lccccccc} 
		\hline
                  Source & Obs. & Filter & Spot No. & Longitude & Co-Latitude & Size & Contrast \\
		 & Date &  &  & $\theta$  (\degr)  & $\phi$ (\degr)  & $\alpha$ (\degr) & $\rho_{\rm{spot}}$ \\
		\hline
		DFOSC & 2015/05/13 & I  & 1 & $ -36.55^{+1.37}_{-1.60}$ & $ 74.93^{+6.45}_{-9.92} $  & $ 9.01^{+ 3.99}_{-3.49} $  & $  0.798^{+0.074}_{-0.126} $  \\
                  \hline
                  DFOSC & 2015/05/17 & I & 1 & $ -14.16^{+1.85}_{-1.87} $  & $ 66.06^{+7.48}_{-4.09} $ & $10.88^{+3.58}_{-4.29} $  & $  0.868^{+0.042}_{-0.069}$    \\
                  DFOSC & 2015/05/17 & I & 2 & $ 24.96^{+1.22}_{-1.13}$  & $ 68.65^{+5.77}_{-5.41}$  & $ 12.22^{+2.76}_{-2.41} $  & $0.873^{+0.041}_{-0.070} $  \\
                  \hline
                  EulerCam & 2011/01/31 & r' & 1 & $ -5.70^{+0.78}_{-0.77}$  & $ 73.34^{+11.66}_{-5.30}$  & $ 6.47^{+1.99}_{-1.07} $ & $ 0.441^{+0.179}_{-0.157} $    \\
                  \hline
		EulerCam & 2011/04/02 & r' & 1 & $ -13.21^{+0.61}_{-0.62} $  & $ 78.43^{+5.56}_{-7.53} $  & $ 10.94^{+1.18}_{-1.21} $  & $ 0.768^{+0.042}_{-0.075}$    \\
                  \hline
                  EulerCam & 2011/05/15 & r' & 1 & $ 14.44^{+3.17}_{-3.08}$  & $  66.67^{+7.28}_{-4.54} $   &  $ 9.69^{+3.17}_{-3.16} $ & $ 0.896^{+0.065}_{-0.133}$    \\
                  \hline
                  TRAPPIST& 2011/04/02 & I+z & 1 & $ -18.49^{+3.27}_{-2.43} $  & $ 72.53^{+5.04}_{-4.87} $  & $  9.93^{+4.00}_{-1.85} $  & $  0.825^{+0.061}_{-0.120} $    \\
	\hline
	\end{tabular}
\end{table*}}

{
\renewcommand{\arraystretch}{1.4}
\begin{table*}[!h]
  \centering
  \small
\caption{Derived spot parameters for the individually analyzed WASP-41b light curves. Parameters are given with their respective one sigma uncertainties.}\label{tab:resspot}
	\begin{tabular}{lccccccc} 
		\hline
                  Source & Obs. & Filter & Spot No. & Longitude & Co-Latitude & Size & Contrast \\
		 & Date &  &  & $\theta$  (\degr)  & $\phi$ (\degr)  & $\alpha$ (\degr) & $\rho_{\rm{spot}}$ \\
		\hline
		DFOSC & 2015/05/13 & I  & 1 & $ -36.87^{+1.35}_{-1.46}$ & $ 75.30^{+10.76}_{-11.26} $  & $ 10.57^{+2.76}_{-2.73}$  & $ 0.795^{+0.080}_{-0.139} $  \\
                  \hline
                  DFOSC & 2015/05/17 & I & 1 & $ -13.60^{+1.38}_{-1.66} $  & $ 67.80^{+7.87}_{-5.54}$ & $ 12.83^{+2.90}_{-3.86}$  & $  0.881^{+0.041}_{-0.079}$  \\
                  DFOSC & 2015/05/17 & I & 2 & $ 23.36^{+1.56}_{-1.49} $  & $ 69.57^{+7.31}_{-6.75}$  & $ 14.60^{+1.83}_{-2.42} $  & $ 0.881^{+0.024}_{-0.054} $  \\
                  \hline
                  EulerCam & 2011/01/31 & r' & 1 & $ -5.77^{+0.75}_{-0.76}$  & $ 77.38^{+8.67}_{-8.91} $  & $ 6.34^{+1.80}_{-0.97}$ & $0.373^{+0.185}_{-0.123} $   \\
                  \hline
		EulerCam & 2011/04/02 & r' & 1 & $ -12.19^{+0.49}_{-0.48} $  & $ 78.10^{+5.59}_{-5.26} $  & $ 11.91^{+1.03}_{-0.71} $  & $ 0.779^{+0.018}_{-0.069} $   \\
                  \hline
                  EulerCam & 2011/05/15 & r' & 1 & $ -13.81^{+2.34}_{-2.61} $  & $ 67.36^{+7.72}_{-4.87} $   &  $ 9.65^{+3.10}_{-3.22} $ & $ 0.873^{+0.076}_{-0.194}$  \\
                  \hline
                  TRAPPIST & 2011/04/02 & I+z & 1 & $ -17.21^{+2.85}_{-3.29} $  & $ 70.88^{+6.32}_{-6.65}$  & $  14.46^{+2.44}_{-3.73}$  & $ 0.851^{+0.053}_{-0.121}$   \\
	\hline
	\end{tabular}
\end{table*}}
\end{appendix}
\end{document}